\documentclass[aps,pre,showpacs,twocolumn]{revtex4}
\usepackage{bm}
\usepackage{graphicx}
\bibstyle{approve.bib}
\usepackage{amssymb}
\usepackage{amsmath}
\usepackage{esint}
\usepackage{epstopdf}
\usepackage{color}
\usepackage{gensymb}
\usepackage{mathtools}
\usepackage{suffix}

\newcommand{\be}{\begin{equation}}
\newcommand{\ee}{\end{equation}}
\newcommand{\bea}{\begin{eqnarray}}
\newcommand{\eea}{\end{eqnarray}}
\newcommand{\lan}{\left\langle}
\newcommand{\ran}{\right\rangle}
\newcommand{\br}{\mathbf{r}}

\newcommand{\bk}{\mathbf{k}}
\newcommand{\bL}{\bar{L}}
\newcommand{\bl}{\mathbf{l}}

\newcommand{\e}{\varepsilon}

\newcommand{\tv}{\tilde{v}}
\newcommand{\tg}{\tilde{G}}

\newcommand{\td}{\tilde{\Delta}}

\newcommand{\tz}{\tilde{z}}
\newcommand{\tL}{\tilde{L}}

\newcommand{\tG}{\tilde{G}}

\newcommand{\bz}{\bar{z}}

\newcommand{\bzp}{\bar{z}_{\rm p}}
\newcommand{\rk}{\bar{k}}
\newcommand{\pa}{\parallel}

\newcommand{\ct}{\cos\te}
\newcommand{\st}{\sin\te}
\newcommand{\te}{\theta_{\rm p}}
\newcommand{\vp}{\varphi_{\rm p}}
\newcommand{\cpk}{\cos\phi_k}

\newcommand{\de}{\partial_z}
\newcommand{\md}{\mathrm{d}}
\newcommand{\mb}{\mathrm{B}}
\newcommand{\p}{{\rm p}}

\newcommand{\m}{{\rm m}}
\newcommand{\w}{{\rm w}}
\newcommand{\rb}{\rho_{\rm b}}
\newcommand{\ef}{{\rm eff}}
\newcommand{\eme}{\varepsilon_{\rm m}}
\newcommand{\ew}{\varepsilon_{\rm w}}
\newcommand{\bph}{\bar{\phi}}
\newcommand{\RP}[1]{\textcolor{black} {#1}}
\newcommand{\SB}[1]{\textcolor{black} {#1}}

\DeclarePairedDelimiterX\MeijerM[3]{\lparen}{\rparen}
{\begin{smallmatrix}#1 \\ #2\end{smallmatrix}\delimsize\vert\,#3}

\newcommand\MeijerG[8][]{G^{\,#2,#3}_{#4,#5}\MeijerM[#1]{#6}{#7}{#8}}

\WithSuffix\newcommand\MeijerG*[7]{G^{\,#1,#2}_{#3,#4}\MeijerM*{#5}{#6}{#7}}

\begin{document}

\title{\RP{Orientational transition and complexation of DNA with anionic membranes:\\
weak and intermediate electrostatic coupling}}

\author{Sahin Buyukdagli$^{1}$\footnote{email:~\texttt{buyukdagli@fen.bilkent.edu.tr}}  
and Rudolf Podgornik$^{2,3,4}$\footnote{email:~\texttt{podgornikrudolf@ucas.ac.cn}}}
\address{$^1$Department of Physics, Bilkent University, Ankara 06800, Turkey\\
$^2$School of Physical Sciences and Kavli Institute for Theoretical Sciences,
University of Chinese Academy of Sciences, Beijing 100049, China\\
$^3$CAS Key Laboratory of Soft Matter Physics, Institute of Physics,
Chinese Academy of Sciences (CAS), Beijing 100190, China\\
$^4$Department of Physics, Faculty of Mathematics and Physics, University of Ljubljana}

\begin{abstract}
We characterize the role of charge correlations in the \RP{adsorption}  of a \RP{short}, rod-like anionic \RP{polyelectrolyte} onto a similarly charged membrane. Our theory reveals two different mechanisms driving the like-charge \RP{polyelectrolyte}-membrane complexation: in weakly charged membranes, repulsive \RP{polyelectrolyte}-membrane interactions lead to the interfacial depletion and a parallel orientation of the \RP{polyelectrolyte}  \RP{with respect to the membrane}; \RP{while in the} intermediate membrane charge regime, the interfacial counterion excess gives rise to an \RP{attractive ''salt- induced'' image force}. 
This furthermore results in an orientational transition from a parallel to a perpendicular configuration and \RP{a subsequent short-ranged} like-charge adsorption of the \RP{polyelectrolyte} to the substrate. 
\RP{A further increase of the membrane charge engenders a charge inversion}, originating from surface-induced ionic correlations, that act as a separate mechanism capable of triggering the like-charge \RP{polyelectrolyte}-membrane complexation over an extended distance interval from the membrane surface. The emerging 
picture of this complexation phenomenon 
identifies the 
\RP{interfacial ''salt- induced'' image forces} as a powerful control mechanism in 
\RP{polyelectrolyte-membrane} complexation.
\end{abstract}


\date{\today}
\maketitle   

\section{Introduction}


Electrostatic interactions play a major role in the regulation of \RP{different biological processes in animate matter}~\cite{biomatter}. The characterization of these interactions is essential for an accurate insight into in vivo biological processes as well as for the optimization of biotechnological methods intending to analyze and manipulate living structures. From gene therapeutic approaches~\cite{Levin1999,PodgornikRev,Molina2013} to nanopore-based biosensing methods~\cite{Tapsarev,Buyukrev},  the 
\RP{details} of various biological processes depend intimately on the nature and strength of the electrostatic coupling between macromolecular charges. Along these lines, the attraction between similarly charged macromolecules has been one of the most fascinating observations in biological physics \cite{burudjerdi,perspective}. In addition to its scientific appeal, the understanding of this seemingly counterintuitive  phenomenon is also \RP{important in order to understand a variety of biological phenomena}, such as the stability of DNA molecules around histones~\cite{PodgornikRev} and anionic membrane assemblies~\cite{Molina2013}, or the condensation in \RP{dense solutions of like-charged polyelectrolytes}, mediated by cationic agents in general ~\cite{Delsanti1994,Raspaud1999,Sabbagh2000}.

The condensation of similarly charged polyelectrolytes has been characterized by intensive theoretical advances that took into account either the one-loop (1l)-level charge fluctuations around the mean-field (MF) Poisson-Boltzmann (PB) electrostatics~\cite{Ha1997,Golestanian1999,Levin1999} \RP{or the non-mean-field states characterized by strong coupling electrostatics \cite{burudjerdi,perspective}}. More recently, the binding of anionic polyelectrolytes onto like-charged membranes has also attracted increasing interest. This partly stems from the high potential of anionic liposomes in gene therapeutic applications \cite{PodgornikRev}; unlike their cationic counterpart of high cytotoxicity, anionic liposome-DNA complexes are efficient gene delivery tools of low toxicity and high transfection efficiency~\cite{Molina2014II}. However, in physiological salt conditions, the stability of these complexes is  weakened by the electrostatic like-charge DNA-liposome repulsion. Thus, the optimization of this genetic manipulation technique requires the identification of the physiological conditions maximizing the cohesion of the DNA with the anionic phospholipid.  This task necessitates in turn a detailed characterization of the mechanism behind the like-charge polyelectrolyte-membrane complexation.

In recent adsorption experiments \cite{Molina2014,Qiu2015,Tiraferri2015,Fries2017} and numerical simulations of DNA molecules at anionic membranes~\cite{Molina2014II,Levin2016},  the like-charge polyelectrolyte-membrane attraction was found to be strongly enhanced by multivalent counterions. Since the electrostatic coupling strength of the system grows with the ion valency, this observation points out ionic correlations as the driving force of the like-charge polyelectrolyte-membrane complexation, \RP{either at intermediate coupling stemming from the fluctuations around the mean-field ground state, or at strong coupling conditions where they are the result of altogether non-mean-field like states} \cite{perspective}. 

 \textcolor{black}{The adsorption of anionic polymers onto cationic substrates has been extensively studied at the MF electrostatic level by functional integral techniques enabling the full consideration of conformational polymer fluctuations~\cite{Podgornik1991,Borukhov1999,Cheng2005,Cherstvy2011} as well as by coarse-grained computer simulations~\cite{Farago2006,Farago2009}. In addition, Nguyen and Shklovskii investigated the alteration of the interaction between two spherical macromolecules upon the adsorption of an oppositely charged polyelectrolyte onto their surface, and the resulting charge inversion of the polymer and/or the polyelectrolyte by this complexation~\cite{Nguyen2001}. Then, in Ref.~\cite{Bohinc2009,Bohinc2012}, an electrostatic MF formalism has been used to show that divalent cations favour the adsorption of DNA molecules onto zwitterionic lipids characterized by a dipolar surface charge distribution.} 
 
 The first theory of like-charge polyelectrolyte-membrane interactions including \SB{charge} correlations was developed by Sens and Joanny for counterion-only Coulomb fluids ~\cite{Sens2000}. By calculating the leading order correlation-correction to the MF PB potential, the Authors showed that the form of the resulting polyelectrolyte self-energy indeed implies an attractive contribution to the polyelectrolyte-membrane coupling. In Ref.~\cite{Buyuk2016}, one of us (SB) introduced a \SB{precise} derivation of the correlation-corrected polyelectrolyte grand potential from the weak-coupling variational grand potential of the system, considering exclusively the parallel and perpendicular configurations of the polyelectrolyte, while the physiological conditions for the like-charge polyelectrolyte-membrane attraction were characterized at finite salt. 

In this work, we generalize the theory of Ref.~\cite{Buyuk2016} in two directions. In Section~\ref{polmod}, we first extend the polyelectrolyte model of Ref.~\cite{Buyuk2016} by introducing an additional angular degree of freedom that enables the rotations of the polyelectrolyte under the effect of its coupling with the liquid and substrate. Then, we generalize the {\sl test charge theory} of Ref.~\cite{Buyuk2016} by carrying out the systematic derivation of the electrostatic polyelectrolyte grand potential directly from the partition function of the system. This results in a polyelectrolyte grand potential that is perturbative in the polyelectrolyte charge, but exact in terms of electrostatic ion-membrane interactions \RP{up to the one-loop fluctuation level}. 

\begin{figure}
\includegraphics[width=1.\linewidth]{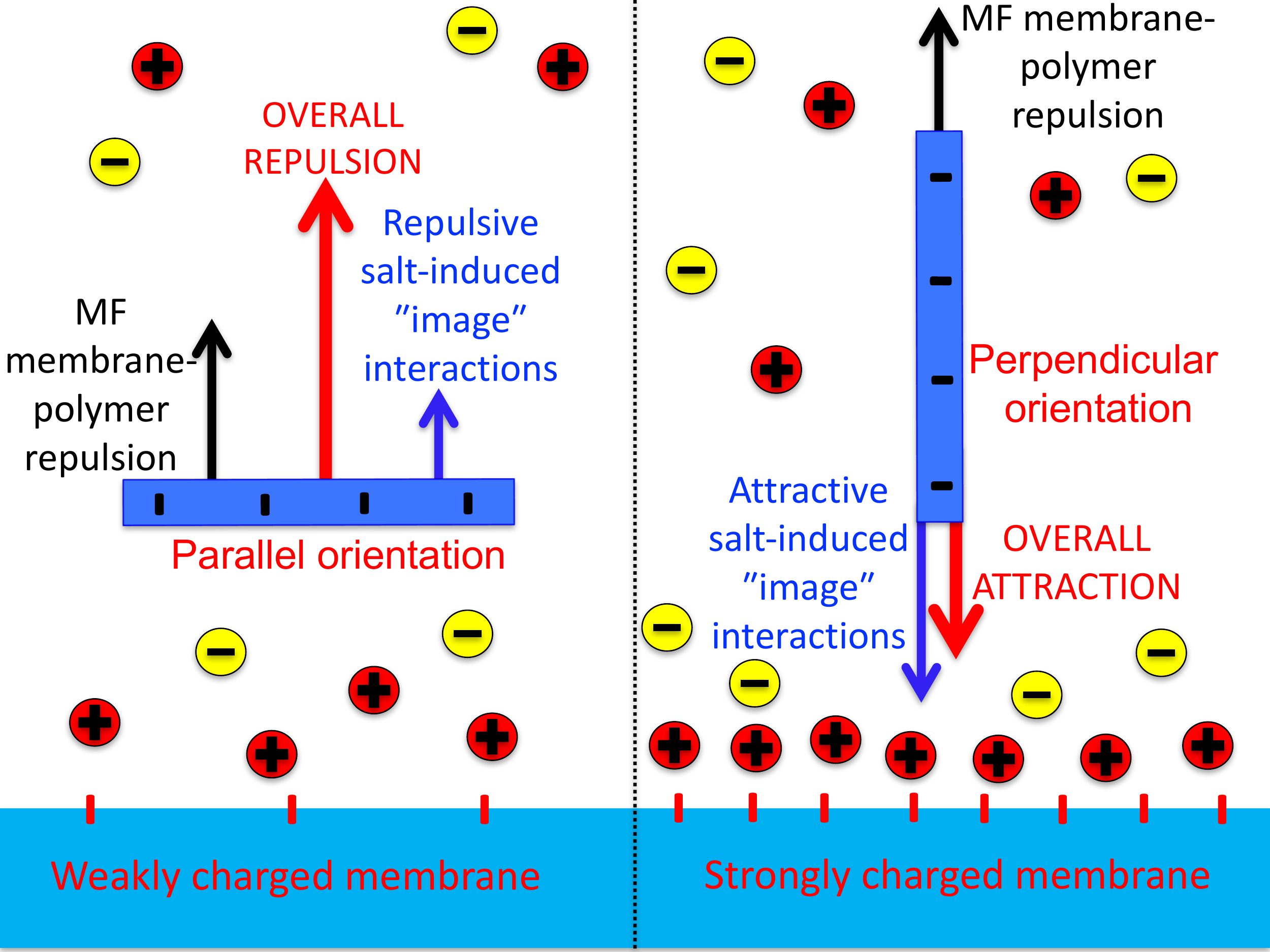}
\caption{(Color online) Schematic depiction of the electrostatic forces acting on the anionic polyelectrolyte close to the similarly charged membrane. In weakly charged membranes, the repulsive MF polyelectrolyte-membrane interaction and the \SB{interfacial ''salt- induced'' image forces driven by charge correlations} lead to the repulsion and the parallel orientation of the polyelectrolyte. In strongly charged membranes, the interfacial counterion excess turns the  \SB{''salt- induced'' image} interactions from repulsive to attractive. This triggers the orientational transition of the polyelectrolyte from the parallel to the perpendicular configuration and the like-charged adsorption of the molecule by the membrane.}.
\label{figsc}
\end{figure}

In Section~\ref{mf}, we characterize polyelectrolyte-membrane interactions in the MF regime of weakly charged membranes in contact with a symmetric monovalent salt solution.  Within the generalized test-charge formalism, Section~\ref{corsalt} deals with the case of weak to intermediate 
membrane charges where the emerging ionic correlations are handled within the 1l theory of inhomogeneous electrolytes. \RP{The weak charge regime would correspond to univalent ions, while the intermediate charge regime would correspond to divalent ions. } Our main findings are summarized in Fig.~\ref{figsc}. The polyelectrolyte-membrane interactions are mainly governed by the charge coupling and the local 
''salt- induced'' image force due to polyelectrolyte charges in an inhomogeneously partitioned electrolyte \cite{saltimage}. In weakly charged membranes, the polyelectrolyte-membrane charge interactions and 
''salt- induced'' image forces of repulsive nature result in the interfacial exclusion of the polyelectrolyte and a parallel orientation of the molecule with respect to the membrane substrate surface. In the intermediate 
membrane charge regime, the counterion excess close to the membrane surface enhances the screening ability of the interfacial electrolyte, and turns the ''salt- induced'' image interaction from repulsive to attractive. Beyond a characteristic membrane charge strength, the attractive ''salt- induced'' image 
interactions take over the repulsive polyelectrolyte-membrane charge coupling, and switch the net force from repulsive to attractive. This leads to the orientational transition of the polyelectrolyte from a parallel to a perpendicular configuration and a consequent adsorption of the molecule by the like-charged membrane. At yet higher membrane charge strengths, correlations give rise to the membrane charge inversion (CI).  The attractive coupling between the polyelectrolyte and the inverted membrane charge acts as a secondary mechanism, inducing the like-charge polyelectrolyte attraction over a larger distance from the membrane surface. Finallly,  for an analytical insight into the effect of the ion multivalency, membrane charge strength, and polyelectrolyte charge and length on the like-charge polyelectrolyte adsorption, we investigate in Section~\ref{counon} polyelectrolyte-membrane interactions in mono- and divalent counterion liquids. In agreement with adsorption experiments~\cite{Molina2014,Qiu2015,Tiraferri2015,Fries2017} and simulations~\cite{Molina2014II,Levin2016}, we find that the presence of multivalent cations enhances the screening ability of the interfacial liquid and strengthens the like-charge polyelectrolyte-membrane complexation. The limitations of our theory and possible extensions are discussed in Conclusions. 

\section{Polyelectrolyte Model and Electrostatic Formalism}
\label{polmod}

\subsection{Charge Composition of the System}
\label{chcom}

\begin{figure}
\includegraphics[width=1.\linewidth]{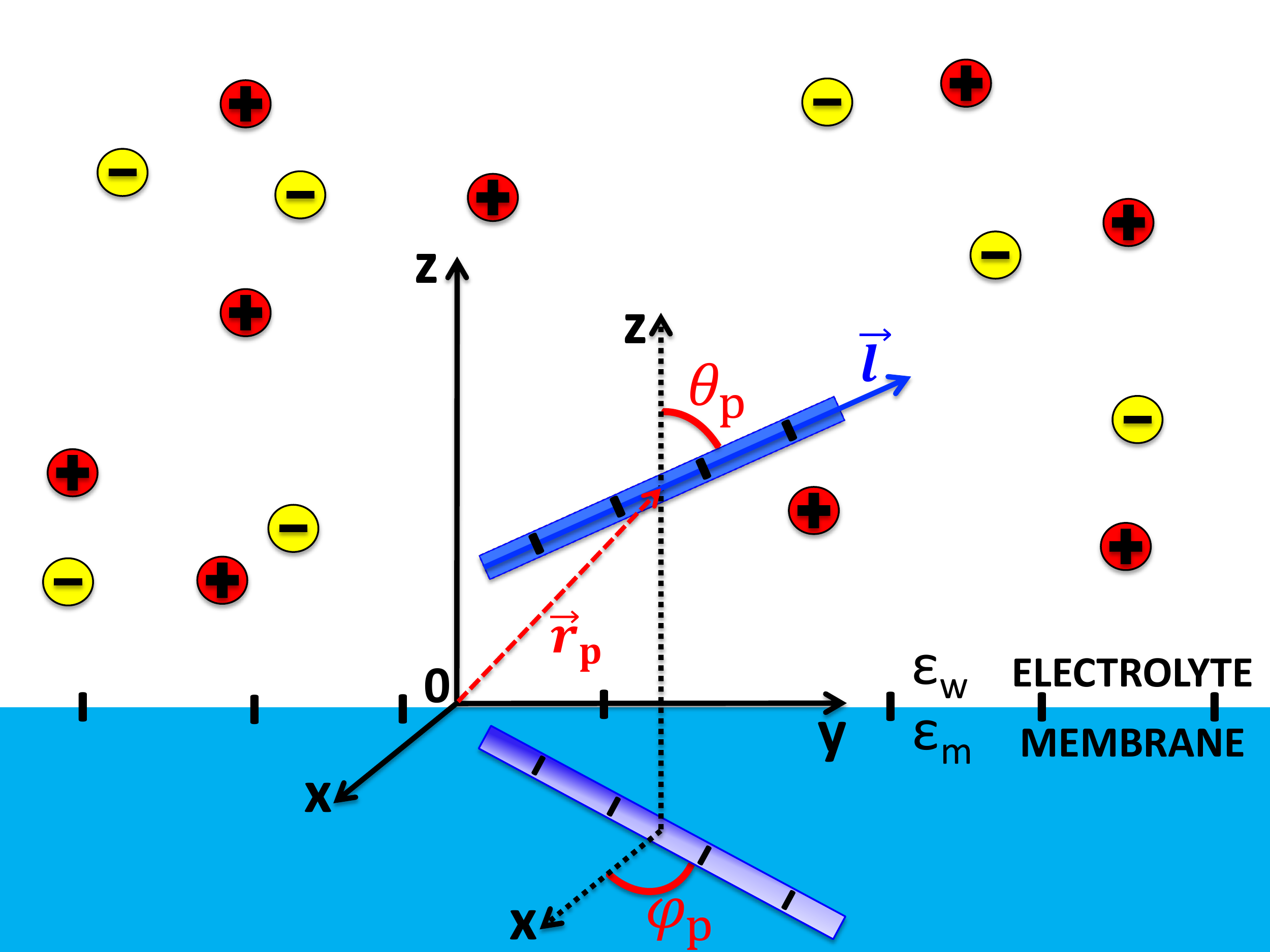}
\caption{(Color online) Schematic depiction of the rotating stiff polyelectrolyte immersed in a charged solution of $p$ ionic species located at $z>0$. The ion species $i$ has valency $q_i$ and bulk concentration $\rho_{{\rm b}i}$. The ion-free membrane at $z<0$ carries an anionic surface charge of density $-\sigma_{\rm m}$. The anionic polyelectrolyte has linear charge density $-\tau$ and length $L$. The corotating coordinate $l$ located on the polyelectrolyte is defined in the interval $-L/2\leq l\leq L/2$. The CM coordinate $\br_\p=(x_\p,y_\p,z_\p)$ is located at $l=0$.}
\label{fig1}
\end{figure}

The schematic depiction of the interacting polyelectrolyte-membrane complex is displayed in Fig~\ref{fig1}. The membrane of dielectric permittivity $\e_{\rm m}$ and negative \SB{interfacial} charge density \SB{$-\sigma_{\rm m}$} is located in the $x-y$ plane and occupies the region $z\leq0$. The electrolyte solution of permittivity $\e_{\rm w}=80$ is located in the half space $z\geq0$. Thus, the dielectric permittivity profile reads
\be
\label{diper}
\e(\br)=\e(z)=\e_{\rm m}\theta_{\rm s}(-z)+\e_{\rm w}\theta_{\rm s}(z),
\ee
\RP{where $\e_{\rm m} = 2$ is the assumed value of the dielectric permittivity of the membrane.} The electrolyte is composed of $p$ ionic species, with the species $i$ having valency $q_i$, fugacity $\Lambda_i$, and bulk concentration $\rho_{bi}$. The polyelectrolyte of length $L$ is a rotating stiff rod of negative line charge density $-\tau$. The latter will be set to the dsDNA value $\tau=2/(3.4\;\mbox{{\AA}})$, unless stated otherwise. Our stiff polyelectrolyte approximation is motivated by the large persistence length $\ell_{\rm p}\approx50$ nm of DNA in monovalent salt at physiological concentrations.

The rotations of the molecule with the center-of-mass (CM) position $\br_\p=(x_\p,y_\p,z_\p)$ are characterized by the polar and azimuthal angles $\te$ and $\vp$. Furthermore, the magnitude of the corotating axis $\bl$ along the polyelectrolyte is defined in the interval $-L/2\leq l\leq L/2$. Thus, the Cartesian coordinates on the polyelectrolyte can be expressed in a parametric form as
\bea
\label{c1}
x\SB{(l)}&=&x_\p+l\st\cos\varphi_\p,\\
\label{c2}
y\SB{(l)}&=&y_\p+l\st\sin\varphi_\p,\\
\label{c3}
z\SB{(l)}&=&z_\p+l\ct.
\eea
Moreover, the steric constraints $z_\p\pm L/2\ct\geq0$ imposed by the hard membrane wall restrict the polyelectrolyte rotations to the interval $\theta_-\leq\te\leq\theta_+$ with the angles
\be\label{thmn}
\theta_-=\arccos\left\{\mathrm{min}\left(1,\frac{2z_\p}{L}\right)\right\},\hspace{3mm}\theta_+=\pi-\theta_-.
\ee

\subsection{Generalized Test-Charge Theory}
\label{tch}

In this part, we extend the weak-coupling test charge theory of Ref.~\cite{Buyuk2016} to the case of 
\RP{intermediate-coupling} charge strength. The grand-canonical partition function of the system can be expressed as a functional integral over a fluctuating electrostatic potential $\phi(\br)$~\cite{Podgornik88}, 
\be\label{zg2}
Z_{\rm G}=\int \mathcal{D}\phi\;e^{-H[\phi]}, 
\ee
with the 
\RP{effective ''field-action'', given by }
\bea\label{HamFunc}
H[\phi]&=&\frac{k_{\rm B}T}{2e^2}\int\mathrm{d}\br\;\e(\br)\left[\nabla\phi(\br)\right]^2-i\int\mathrm{d}\br\sigma(\br)\phi(\br)\nonumber\\
&&-\sum_{i=1}^p\Lambda_i \int\mathrm{d}\br\;e^{iq_i\phi(\br)}\theta_{\rm s}(z).
\eea
The first term of Eq.~(\ref{HamFunc}) corresponding to the free energy of the solvent includes the Boltzmann constant $k_{\rm B}$,  the liquid temperature $T=300$ K, and the electron charge $e$. The second term takes into account the total macromolecular charge density distribution
\be\label{mc}
\sigma(\br)=\sigma_{\rm m}(\br)+\sigma_\p(\br),
\ee
\SB{where the membrane and polyelectrolyte charge density functions are respectively given by
\bea
\sigma_{\rm m}(\br)&=&-\sigma_{\rm m}\delta(z),\\
\sigma_{\rm p}(\br)&=&-\tau\int_{-L/2}^{L/2}\mathrm{d}l\;\delta\left[\br-\br(l)\right],
\eea
with the vector $\br(l)=x(l)\hat{u}_x+y(l)\hat{u}_y+z(l)\hat{u}_z$.} Finally, the third term of Eq.~(\ref{HamFunc}) corresponds to the fluctuating density of mobile ions.

The rotating polyelectrolyte obviously breaks the planar symmetry of the system, \RP{rendering an explicit analytical solution unreachable. The strategy of the test charge theory then consists of reintroducing the simplifying planar symmetry at the price of treating the polyelectrolyte as a small perturbation}. \RP{Following this approach and Taylor expanding the partition function~(\ref{zg2}) to the quadratic order in the polyelectrolyte charge $\sigma_\p(\br)$, one remains with}
\bea\label{zg3}
Z_{\rm G}&=&Z_0\left\{1+i\int\mathrm{d}\br\sigma_\p(\br)\lan\phi(\br)\ran_0\right.\\
&&\left.\hspace{7mm}-\frac{1}{2}\int\mathrm{d}\br\mathrm{d}\br'\sigma_\p(\br)\lan\phi(\br)\phi(\br')\ran_0\sigma_\p(\br')\right\},\nonumber
\eea
where we defined the polyelectrolyte-free partition function 
\be\label{z0}
Z_0=\int \mathcal{D}\phi\;e^{-H_0[\phi]},
\ee
with the corresponding Hamiltonian functional
\bea\label{H0}
H_0[\phi]&=&\frac{k_{\rm B}T}{2e^2}\int\mathrm{d}\br\;\e(\br)\left[\nabla\phi(\br)\right]^2-i\int\mathrm{d}\br\sigma_{\rm m}(\br)\phi(\br)\nonumber\\
&&-\sum_{i=1}^p\Lambda_i \int\mathrm{d}\br\;e^{iq_i\phi(\br)}\theta_{\rm s}(z).
\eea
In Eq.~(\ref{zg3}), the bracket is defined as the field theoretic average with the polyelectrolyte-free Hamiltonian, i.e.
\be
\label{br}
\lan F[\phi]\ran_0=\frac{1}{Z_0}\int \mathcal{D}\phi\;e^{-H_0[\phi]}F[\phi].
\ee

At the same quadratic order in the polyelectrolyte charge $\sigma_\p(\br)$, the dimensionless electrostatic grand potential $\beta\Omega_{\rm G}\equiv-\ln Z_{\rm G}$ follows as
\bea\label{og}
\beta\Omega_{\rm G}&=&\beta\Omega_0+\int\mathrm{d}\br\sigma_\p(\br)\bph(\br)\\
&&+\frac{1}{2}\int\mathrm{d}\br\mathrm{d}\br'\sigma_\p(\br)G(\br,\br')\sigma_\p(\br'),\nonumber
\eea
where we defined the polyelectrolyte-free grand potential $\beta\Omega_0=-\ln Z_0$, and the real average potential and two-point correlation function of the fluctuating potential $\phi(\br)$,
\bea
\label{pm}
\bph(\br)&=&-i\lan\phi(\br)\ran_0,\\
\label{gr}
G(\br,\br')&=&\lan\phi(\br)\phi(\br')\ran_0-\lan\phi(\br)\ran_0\lan\phi(\br')\ran_0.
\eea
From Eq.~(\ref{og}), the polyelectrolyte grand potential defined as $\Omega_\p=\Omega_{\rm G}-\Omega_0$ follows in the form
\be\label{ogp}
\beta\Omega_\p=\int\mathrm{d}\br\sigma_\p(\br)\bph(\br)+\frac{1}{2}\int\mathrm{d}\br\mathrm{d}\br'\sigma_\p(\br)G(\br,\br')\sigma_\p(\br').
\ee
By subtracting from the grand potential~(\ref{ogp}) its bulk limit, one gets the renormalized polyelectrolyte grand potential 
\be\label{grren}
\Delta\Omega_\p=\Omega_{\rm pm}+\Delta\Omega_{\rm pp},
\ee
with the direct coupling energy between the polyelectrolyte and the membrane charges 
\be\label{ogpm}
\beta\Omega_{\rm pm}=\int\mathrm{d}\br\sigma_\p(\br)\bph(\br),
\ee
and the polyelectrolyte self energy  renormalized by its bulk value
\be
\label{ogpp}
\beta\Delta\Omega_{\rm pp}=\frac{1}{2}\int\mathrm{d}\br\mathrm{d}\br'\sigma_\p(\br)\left[G(\br,\br')-G_{\rm b}(\br-\br')\right]\sigma_\p(\br').
\ee
In Eq.~(\ref{ogpp}), the correlation function $G(\br,\br')$ corresponds to the potential induced by a point charge at $\br'$ at the point $\br$. Moreover, the bulk correlator $G_{\rm b}(\br-\br')$ is the limit of this correlation function in the ionic reservoir located infinitely far from the membrane. \SB{We finally note that because the polymer-membrane interaction energy $\Omega_{\rm pm}$ vanishing in the bulk does not have to be renormalized, its symbolic notation $\Omega_{\rm pm}$ is not preceeded by the symbol $\Delta$.}

The grand potential~(\ref{grren}) corresponds to the adiabatic work required for bringing the polyelectrolyte from the bulk reservoir to the distance $z_\p$ from the membrane. It is important to note that within the test charge approach, the potential $\bph(\br)$ in the coupling energy~(\ref{ogpm}) originates solely from the charged membrane and it is screened exclusively by the mobile ions. Thus, the potentials $\bph(\br)$ and $G(\br,\br')$ \RP{lack to the lowest order any contribution from the presence of the polyelectrolyte charges}. Finally, Eq.~(\ref{ogpp}) corresponds to the polyelectrolyte self-energy dressed by the electrolyte-membrane interactions. In Section~\ref{corsalt}, we show that this self-energy driven purely by correlations vanishes in the MF regime. 

We emphasize that the derivation of the formula~(\ref{grren}) did not involve any assumption on the strength of the electrostatic coupling between the mobile ions and the charged membrane. Thus, by calculating the average potential $\bph(\br)$ and the Green's function $G(\br,\br')$ at the appropriate approximation level, Eq.~(\ref{grren})  allows to evaluate the polyelectrolyte grand potential from the weak to the strong electrolyte-membrane coupling regime. \RP{In the present work, we will consider exclusively the weak  coupling regime, valid for monovalent ions, and the intermediate coupling regime, valid for divalent cations}. The strong coupling regime of higher ionic valencies will be considered in an upcoming work. We finally note that as the test-charge approach is based on the Taylor expansion of the grand potential in terms of the polyelectrolyte charge $\sigma_\p(\br)$, our theory treats the polyelectrolyte-membrane interactions at the weak coupling (WC) level. This approximation is based on the superposition principle where the additivity of the average membrane and rod potentials is assumed. 

\subsection{Introducing the Plane Symmetry}
\label{ps}

The form of the grand potential components~(\ref{ogpm}) and~(\ref{ogpp}) can be simplified by accounting for the planar symmetry implying $\bph(\br)=\bph(z)$ and $G(\br,\br')=G(\br_{\pa}-\br'_{\pa},z,z')$. Based on the latter equality, we Fourier-expand the Green's function as 
\be\label{6}
G(\br,\br')=\int\frac{d^2\bk}{4\pi^2}e^{i\bk\cdot\left(\br_{\pa}-\br'_{\pa}\right)}\tg(z,z';k).
\ee
In order to simplify the notation, from now on, the dependence of the potentials and auxiliary functions on the wave vector $\bk$ will be omitted. Using in Eqs.~(\ref{ogpm}) and~(\ref{ogpp}) the Fourier expansion~(\ref{6}) and the coordinates~(\ref{c1})-(\ref{c3}), the grand potential components become
\bea\label{9}
\beta\Omega_{\rm pm}(z_\p,\te)&=&-\tau\int_{-L/2}^{L/2}\md l\;\bph\left(z_\p+l\ct\right),\\
\label{9II}
\beta\Delta\Omega_{\rm pp}(z_\p,\te)&=&\frac{\tau^2}{2}\int\frac{\md\bk}{4\pi^2}\int_{-L/2}^{L/2}\md l\int_{-L/2}^{L/2}\md l'e^{i\bk\cdot(\bl-\bl')}\nonumber\\
&&\hspace{.5cm}\times\delta\tG\left(z_\p+l\ct,z_\p+l'\ct\right),\nonumber\\
\eea
with the infinitesimal wave vector $\md\bk=\md k_x\md k_y=k\md k\md\phi_k$, the scalar product $\bk\cdot\bl=kl\st\cpk$, and the renormalized Green's function
\be\label{rengr}
\delta\tG\left(z_1,z_2\right)=\tG\left(z_1,z_2\right)-\tG_{\rm b}\left(z_1-z_2\right). 
\ee

The orientation-averaged polyelectrolyte number density is defined in terms of the polyelectrolyte grand potential~(\ref{grren}) as
\be\label{dn}
\rho_{\rm p}(z)=\frac{\rho_{\rm pb}}{2}\int_{\theta_-}^{\theta_+}\mathrm{d}\theta\sin\theta e^{-\beta\Delta\Omega_\p(z_\p,\te)}
\ee
where $\rho_{\rm bp}$ is the bulk polyelectrolyte concentration. Moreover, the average orientation of the polyelectrolyte can be quantified in terms of the \RP{(nematic) orientational order parameter}
\be\label{or}
S_\p(z_\p)=\frac{3}{2}\left[\lan\cos^2\theta_\p\ran-\frac{1}{3}\right],
\ee
where we introduced the orientational average  
\be\label{defi}
\lan f(\theta_\p)\ran=\frac{\int_{\theta_-}^{\theta_+}\mathrm{d}\theta_\p\sin\theta_\p f(\theta_\p) e^{-\beta\Delta\Omega_\p(z_\p,\te)}}{\int_{\theta_-}^{\theta_+}\mathrm{d}\theta_\p\sin\theta_\p e^{-\beta\Delta\Omega_\p(z_\p,\te)}}.
\ee
Eq.~(\ref{or}) yields $S_\p(z_{\rm p})=-1/2$ for the exact parallel polyelectrolyte orientation with the membrane surface and $S_\p(z_{\rm p})=1$ for the strictly perpendicular orientation. These two regimes are separated by the freely rotating dipole limit $S_\p(z_{\rm p})=0$ reached for vanishing electrostatic and steric polyelectrolyte-membrane interactions, i.e. for $\Delta\Omega_\p(z_\p,\te)=0$,  $\theta_-=0$, and $\theta_+=\pi$.  

In order to illustrate the effect of the steric penalty, we consider the simplest non-trivial case of a neutral polyelectrolyte where electrostatic polyelectrolyte-membrane interactions vanish. In this case, the polyelectrolyte density~(\ref{dn}) and orientational order parameter~(\ref{or}) become
\bea\label{dest}
\rho_{\rm p}(z_\p)&=&\rho_{\rm pb}\hspace{1mm}\mathrm{min}\left(1,\frac{2z_\p}{L}\right),\\
\label{orst}
S_\p(z_\p)&=&\frac{1}{2}\hspace{1mm}\mathrm{min}\left(0,\frac{4z_\p^2}{L^2}-1\right).
\eea
Eqs.~(\ref{dest}) and~(\ref{orst}) reported in Figs.~\ref{fig3}(a) and (b) by the dotted curves indicate that for $z_\p<L/2$, the steric repulsion by the membrane results in the polyelectrolyte depletion $\rho_{\rm p}(z)<\rho_{\rm pb}$, and also the parallel alignment of the molecule with the membrane surface, i.e. $S_\p(z_{\rm p})<0$. In the region $z_\p>L/2$ where the steric effect vanishes, one recovers the bulk behavior  $\rho_{\rm p}(z_\p)=\rho_{\rm pb}$ and $S_\p(z_{\rm p})=0$.

\subsection{One-Loop Formalism of Electrostatic Interactions}

In this work, we consider polyelectrolyte-membrane interactions solely in the regimes of weak to intermediate coupling, \RP{valid for monovalent and divalent ions, basing our approach on the 1l fluctuation theory of Refs.~\cite{Netz2000,Buyuk2012}}. Thus, the mean value and correlator of the fluctuating potential in Eqs.~(\ref{ogpm})-(\ref{9II}) will be approximated by their 1l-level counterpart $\phi_{\rm m}(z)$ and $v(\br,\br')$, i.e.
\bea
\label{mp1}
\bph(z)&=&\phi_{\rm m}(z),\\
\label{v1}
G(\br,\br')&=&v(\br,\br').
\eea

Within the 1l approximation, the average potential $\phi_{\rm m}(z)$ in Eqs.~(\ref{9}) and~(\ref{mp1}) is given by the superposition of the MF potential $\phi^{(0)}_{\rm m}(z)$ and the 1l correction $\phi^{(1)}_{\rm m}(z)$ including the leading order charge correlations~\cite{Netz2000},
\be\label{sup}
\phi_{\rm m}(z)=\phi^{(0)}_{\rm m}(z)+\phi^{(1)}_{\rm m}(z).
\ee
Taking also into account the 1l limit of the self-energy $\Delta\Omega^{(1)}_{\rm pp}$ that will be obtained below from Eq.~(\ref{9II}), the 1l-level polyelectrolyte grand potential~(\ref{grren}) becomes
\be\label{gr1l}
\Delta\Omega_\p(z_\p,\te)=\Omega^{(0)}_{\rm pm}(z_\p,\te)+\Omega^{(1)}_{\rm pm}(z_\p,\te)+\Delta\Omega^{(1)}_{\rm pp}(z_\p,\te).
\ee

In Eq.~(\ref{gr1l}), the MF and 1l components of the polyelectrolyte-membrane coupling potential~(\ref{9}) are
\be\label{pm1l}
\beta\Omega^{(i)}_{\rm pm}(z_\p,\te)=-\tau\int_{-L/2}^{L/2}\md l\;\phi^{(i)}_{m}\left(z_\p+l\ct\right)
\ee
for $i=0$ and $1$. The  MF potential $\phi^{(0)}_{\rm m}(z)$ in Eq.~(\ref{pm1l}) with $i=0$ solves the PB equation 
\be\label{4}
\frac{k_{\rm B}T}{e^2}\partial_z\e(z)\partial_z\phi^{(0)}_{\rm m}(z)+\sum_{i=1}^pq_in_i(z)=\sigma_{\rm m}\delta(z),
\ee
where we introduced the MF-level ion number density  
\be\label{ni}
n_i(z)=\rho_{{\rm b}i}\theta_{\rm s}(z)e^{-q_i\phi^{(0)}_{\rm m}(z)}.
\ee
Then, the 1l-level Green's function in Eqs.~(\ref{9II}) and~(\ref{v1}) solves the kernel equation
\be
\label{5}
\frac{k_{\rm B}T}{e^2}\nabla\e(\br)\cdot\nabla v(\br,\br')-\sum_{i=1}^pq_i^2n_i(z)v(\br,\br')=-\delta(\br-\br').
\ee
Using the Fourier expansion~(\ref{6}), Eq.~(\ref{5}) simplifies to
\bea
\label{7}
\left[\de\e(z)\de-\e(z)p^2(z)\right]\tv(z,z')=-\frac{e^2}{k_{\rm B}T}\delta(z-z'),
\eea
with the local screening function
\be
\label{ep}
p^2(z)=k^2+\frac{e^2}{\e(z)k_{\rm B}T}\sum_{i=1}^pq_i^2n_i(z).
\ee

In the single interface system of Fig.~\ref{fig1}, the general solution to Eq.~(\ref{7}) reads~\cite{Buyuk2012}
\be
\label{8II}
\tv(z,z')=4\pi\ell_{\rm B}\frac{h_+(z_<)h_-(z_>)+\Delta h_-(z_<)h_-(z_>)}{h'_+(z')h_-(z')-h'_-(z')h_+(z')},
\ee
where the functions $h_\pm(z)$ are the homogeneous solutions of Eq.~(\ref{7}),
\be\label{eh}
\left[\partial_z^2-p^2(z)\right]h_{\pm}(z)=0.
\ee
In Eq.~(\ref{8II}), we introduced the auxiliary variables $z_<=\mathrm{min}(z,z')$ and $z_>=\mathrm{max}(z,z')$, and the  function
\be
\label{8III}
\Delta=\frac{h'_+(0)-\eta kh_+(0)}{\eta kh_-(0)-h'_-(0)},
\ee
where we defined the dielectric contrast parameter 
\be\label{con}
\eta=\frac{\e_{\rm m}}{\e_{\rm w}}.
\ee

Finally, the 1l potential correction in Eq.~(\ref{pm1l}) satisfies the differential equation
\be\label{1lp}
\frac{k_{\rm B}T}{e^2}\partial_z\e(z)\partial_z\phi^{(1)}_{\rm m}(z)-\sum_{i=1}^pq_i^2n_i(z)\phi^{(1)}_{\rm m}(z)=-\delta\sigma(z),
\ee
with the non-uniform charge excess 
\be
\label{chex}
\delta\sigma(z)=-\frac{1}{2}\sum_{i=1}^pq_i^3n_i(z)\delta v(z)
\ee
where we introduced the ionic self-energy corresponding to the equal point Green's function renormalized by its bulk limit,
\be\label{is}
\delta v(z)=\int\frac{d^2\bk}{4\pi^2}\left[\tv(z,z)-\lim_{z\to\infty}\tv(z,z)\right].
\ee

\RP{This self-energy~(\ref{is}) embodies two different effects, both rationalizable in terms of image interactions: the first one is the effect of standard dielectric image interactions, pending on the presence of dielectric inhomogeneities in the system; the other one describes the ''salt-induced'' image effects, which are not due to dielectric inhomogeneities but due to an inhomogeneous distribution of the salt in the system, as it is excluded from the membrane phase \cite{saltimage,Buyuk2012}}. 

By using now the kernel Eq.~(\ref{7}) together with the definition of the \SB{inverse operator}
\be
\label{defgr}
\int\mathrm{d}\br''v^{-1}(\br,\br'')v(\br'',\br')=\delta(\br-\br'),
\ee
Eq.~(\ref{1lp}) can be inverted as
\be\label{1lp2}
\phi^{(1)}_{\rm m}(z)=\int_0^\infty\mathrm{d}z'\tv(z,z';k=0)\delta\sigma(z').
\ee

At this point, we wish to emphasize the meaning of the 1l potential correction in Eq.~(\ref{1lp2}). To this end, we first note that in the second term of the PB Eq.~(\ref{4}) taking into account the non-uniform charge screening of the average electrostatic potential, the exponential ion density function $n_i(z)$ includes exclusively the coupling of the mobile charge $q_i$ to the MF average potential $\phi_m^{(0)}(z)$ (see Eq.~(\ref{ni})). According to Eqs.~(\ref{chex}) and~(\ref{1lp2}), the 1l potential correction $\phi^{(1)}_{\rm m}(z)$ accounts for the additional effect of the self-energy $\delta v(z)$ on the mobile ions, and the resulting modification of the MF-level charge screening of the average electrostatic potential.

\begin{table}[ht]
\caption{Electrostatic Model Parameters}
\begin{tabular}{c c}
\hline\hline 
Bjerrum length & $\ell_{\rm B}=\frac{e^2}{4\pi\ew k_{\rm B}T}\approx 7$ {\AA} \\ [1.0 ex] 
\hline
Gouy-Chapman length & $\mu=1/(2\pi q\ell_{\rm B}\sigma_{\rm m})$\\ [1.0 ex] 
\hline
Debye-H\" uckel  screening parameter & $\kappa=\sqrt{8\pi q^2\ell_{\rm B}\rho_{\rm b}}$\\ [1.0 ex] 
\hline
Relative screening strength & $s=\kappa\mu$\\ [1.0 ex] 
\hline
\SB{Auxiliary screening parameter} & \SB{$\gamma=\sqrt{s^2+1}-s$}\\ [1.0 ex] 
\hline
Counterion coupling strength & $\Xi_{\rm c}=\frac{q^2\ell_{\rm B}}{\mu}$\\ [1.0 ex]
\hline
Bulk coupling strength & $\Gamma_{\rm s}=q^2\kappa\ell_{\rm B}=s\;\Xi_{\rm c}$\\ [1.0 ex]
\hline
\end{tabular}
\label{table:nonlin}
\end{table}

\section{Symmetric Monovalent Electrolyte: Mean Field} 
\label{mf}

We investigate here the mean-field theory of polyelectrolyte-membrane interactions in a symmetric 1:1 electrolyte with the ionic valencies $q_i=q=\pm1$ and bulk concentrations $\rho_{{\rm b}i}=\rho_{\rm b}$. Our analysis will be thus limited to weakly charged membranes where ion correlations are negligible. 

\SB{We note that within this MF approach, the ionic fugacities in Eq.~(\ref{HamFunc}) are related to the bulk concentrations as $\Lambda_i=\rho_{{\rm b}i}=\rho_{\rm b}$.} The MF membrane potential solving Eq.~(\ref{4}) reads~\cite{Isr}
\be\label{10}
\phi^{(0)}_{\rm m}(z)=-2\ln\left[\frac{1+\gamma e^{-\kappa z}}{1-\gamma e^{-\kappa z}}\right],
\ee
with the auxiliary parameter 
\be\label{gm}
\gamma=\sqrt{s^2+1}-s.
\ee
In Eq.~(\ref{gm}), we used the dimensionless constant 
\be\label{s}
s=\kappa\mu.
\ee
Eq.~(\ref{s}) includes the Debye-H\"{u}ckel (DH) screening parameter $\kappa$ and Gouy-Chapman (GC) length $\mu$,
\be
\kappa=\sqrt{8\pi q^2\ell_{\rm B}\rho_{\rm b}}\;;\hspace{1cm}\mu=\frac{1}{2\pi q\ell_{\rm B}\sigma_{\rm m}},
\ee
with the Bjerrum length $\ell_{\rm B}=e^2/(4\pi\e_\w k_{\rm B}T)\approx 7$ {\AA} corresponding to the separation distance where two ions interact with thermal energy $k_{\rm B}T$. The DH length $\kappa^{-1}$ corresponds in turn to the characteristic radius of the ionic cloud around a central ion in the bulk region. Finally, the GC length $\mu$ is the thickness of the counterion layer at the membrane surface. Thus, the parameter $s$ in Eq.~(\ref{s}) quantifies the relative density and screening ability of the bulk salt and the interfacial counterions. These definitions are summarized in Table I.

\begin{figure}
\includegraphics[width=1.\linewidth]{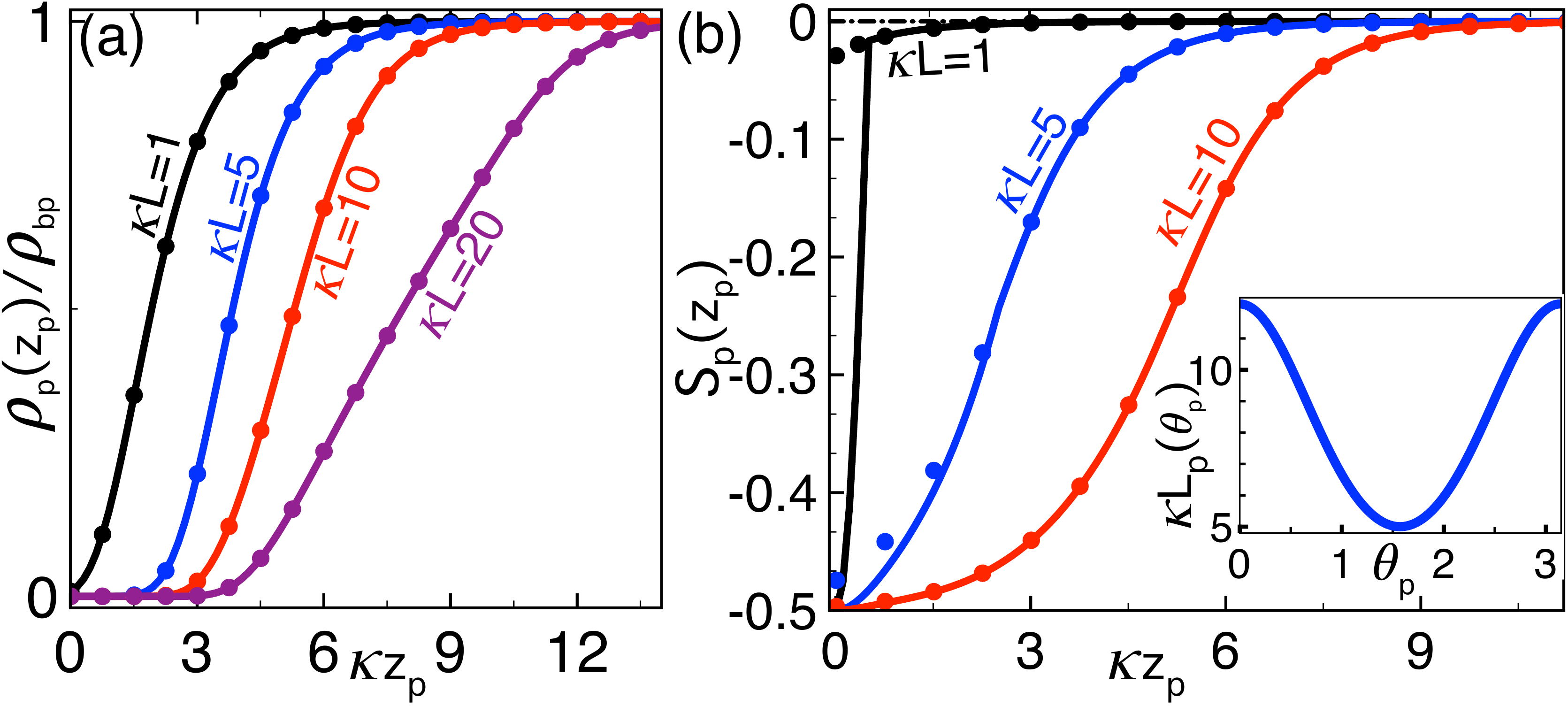}
\caption{(Color online) (a) polyelectrolyte density~(\ref{dn}) and (b) orientational order parameter~(\ref{or}) including the steric rotational penalty (solid curves) and neglecting the steric penalty (dots) at various polyelectrolyte lengths. The inset in (b) displays the variation of the polyelectrolyte grand potential~(\ref{11}) with the polyelectrolyte angle $\theta_p$ in terms of the effective polyelectrolyte length~(\ref{13}). Salt concentration is $\rb=0.1$ M and the membrane charge density $\sigma_{\rm m}=0.1$ $e/\mbox{nm}^2$.}
\label{fig2}
\end{figure}

Substituting now the potential~(\ref{10}) into Eq.~(\ref{pm1l}), the MF polyelectrolyte-membrane interaction energy follows as
\bea
\label{11}
\beta\Omega^{(0)}_{\rm pm}(\tz_\p,\te)&=&\frac{2\tau}{\kappa\ct}\left\{\mathrm{Li}_2\left[\gamma e^{-\tz_-}\right]-\mathrm{Li}_2\left[-\gamma e^{-\tz_-}\right]\right.\\
&&\hspace{1.4cm}\left.-\mathrm{Li}_2\left[\gamma e^{-\tz_+}\right]+\mathrm{Li}_2\left[-\gamma e^{-\tz_+}\right]\right\}.\nonumber
\eea
Eq.~(\ref{11}) includes the polylog function $\mathrm{Li}_2(x)$~\cite{math} and the distance of the polyelectrolyte edges from the membrane,
\be\label{11II}
\tz_\pm=\tz_\p\pm\frac{\tL}{2}\ct,
\ee
with the dimensionless polyelectrolyte distance $\tz_\p=\kappa z_\p$ and length $\tL=\kappa L$. Fig.~\ref{fig2}(a)  displays the MF-level polyelectrolyte density profiles obtained from Eq.~(\ref{dn}) and~(\ref{11}), i.e. by neglecting the 1l grand potential corrections in Eq.~(\ref{gr1l}). The plot shows polyelectrolyte depletion from the vicinity of the membrane surface. Comparison of the results including the steric rotational penalty (solid curves) and without the penalty (dots) indicates that the polyelectrolyte depletion is mainly driven by the electrostatic polyelectrolyte-membrane repulsion and the steric barrier does not bring a relevant contribution. This stems from the fact that for polyelectrolytes of length $\tL\gtrsim1$, the electrostatic polyelectrolyte repulsion occurring on the interval $z_{\rm p}\lesssim L$ is too strong for the steric repulsion at $z_{\rm p}\leq L/2$ to be noticeable.

Due to the salt screening of these repulsive electrostatic interactions, the polyelectrolyte density quickly rises with the distance $z_p$ to its bulk value. Indeed, in the MF DH regime of weak membrane charges where $s\gg1$, one finds that salt screening results in the exponential decay of the MF potential~(\ref{11}),
\be\label{12}
\beta\Omega^{(0)}_{\rm pm}(\tz_\p,\te)\approx\frac{2}{s} \tau L_\p(\te)e^{-\tz_\p},
\ee
where we introduced the effective polyelectrolyte length 
\be
\label{13}
L_\p(\te)=\frac{2\sinh\left(\tL\ct/2\right)}{\kappa\ct}.
\ee 
Fig.~\ref{fig2}(a) also shows that the interfacial polyelectrolyte exclusion layer expands with the length of the molecule,  i.e. $L\uparrow\rho_{\rm p}(z_{\rm p}) \downarrow$ at fixed distance $z_{\rm p}$. According to Eqs.~(\ref{12}) and~(\ref{13}), this results from the intensification of the repulsive polyelectrolyte-membrane coupling with the increase of the polyelectrolyte length, i.e. $L\uparrow\;\Omega^{(0)}_{\rm pm}(\tz_\p,\te)\uparrow$. 

In the inset of Fig.~\ref{fig2}(b), the variation of the polyelectrolyte-membrane interaction energy with the orientational angle $\theta_p$ is illustrated in terms of the effective length~(\ref{13}).  One sees that due to repulsive polyelectrolyte-membrane interactions, the parallel polyelectrolyte orientation $\theta_p=\pi/2$ minimizing the electrostatic interaction energy is the stable polyelectrolyte configuration. This point is also illustrated in the main plot where the order parameter~(\ref{or}) indicates parallel alignment  close to the membrane, i.e. $S_\p(z_\p)\to-0.5$ as $z_\p\to0$. The comparison of the solid curves and dots indicates that the alignment is essentially induced by electrostatic interactions, and the steric penalty plays a noticeable role only close to the membrane surface or for short polyelectrolytes with length $L\sim\kappa^{-1}$. Moving away from the surface, salt screening leads to the gradual loss of the orientational order and the order parameter approaches from below the bulk value $S_\p(z_\p)=0$ indicating free polyelectrolyte rotation.  We finally note that in Fig.~\ref{fig2}(b), the tendency of the polyelectrolyte to orient itself along the membrane increases with its length, i.e. $L\uparrow S_\p(z_{\rm p})\downarrow$. This stems again from the enhancement of the polyelectrolyte-membrane repulsion with the polyelectrolyte length.

\section{Symmetric monovalent electrolyte: 1l correlations}
\label{corsalt}

In this part, we extend the MF analysis of the previous section on weakly charged membranes to the case of strong membrane charges where electrostatic correlations become relevant. To this end, we take into account the 1l-level correlation potentials $\Delta\Omega^{(1)}_{\rm pp}$ and $\Omega^{(1)}_{\rm pm}$ in Eq.~(\ref{gr1l}).

\subsection{Computation of 1l Correction Potentials $\Delta\Omega^{(1)}_{\rm pp}$ and $\Omega^{(1)}_{\rm pm}$}

For the computation of the 1l correction potentials defined in Eqs.~(\ref{9II}) and~(\ref{pm1l}), we review the calculation of the Green's function $v(\br,\br')$ derived in Ref.~\cite{Buyuk2012}. Inserting the MF potential~(\ref{10}) into Eqs.~(\ref{ni}) and~(\ref{ep}), the differential equation~(\ref{eh}) becomes
\be\label{14}
h''_\pm(z)-\left\{p^2+\frac{2\kappa^2}{\sinh^2\left[\kappa(z+z_0)\right]}\right\}h_\pm(z)=0,
\ee
where we introduced the parameter $p=\sqrt{k^2+\kappa^2}$ and the characteristic thickness of the interfacial counterion layer $z_0=\ln(\gamma^{-1})/\kappa$. In Ref.~\cite{Lau}, the solution of Eq.~(\ref{14}) was found as
\be\label{15}
h_\pm(z)=e^{\pm pz}\left\{1\mp\frac{\kappa}{p}\coth\left[\kappa(z+z_0)\right]\right\}.
\ee
With the homogeneous solutions in Eq.~(\ref{15}), the Fourier-transformed Green's function~(\ref{8II}) simplifies to
\be\label{16}
\tv(z,z')=\frac{2\pi\ell_{\rm B}p}{k^2}\left[h_+(z_<)+\Delta h_-(z_<)\right]h_-(z_>)
\ee
where the delta function defined in Eq.~(\ref{8III}) reads
\be\label{17}
\Delta=\frac{\kappa^2\mathrm{csch}^2\left(\kappa z_0\right)+(p_{\rm b}-\eta k)\left[p_{\rm b}-\kappa\coth\left(\kappa z_0\right)\right]}
{\kappa^2\mathrm{csch}^2\left(\kappa z_0\right)+(p_{\rm b}+\eta k)\left[p_{\rm b}+\kappa\coth\left(\kappa z_0\right)\right]}.
\ee
In the bulk limit $z\to\infty$ and $z'\to\infty$, the Fourier-transformed Green's function~(\ref{16}) becomes 
\be\label{17II}
\tv(z,z')\to\tv_{\rm b}(z-z')=\frac{2\pi\ell_{\rm B}}{p_{\rm b}}e^{-|z-z'|}. 
\ee
Thus, the bulk Green's function follows from Eq.~(\ref{6}) as the screened Coulomb potential
\be
\label{18}
v_{\rm b}(\br-\br')=\ell_{\rm B}\frac{e^{-\kappa|\br-\br'|}}{|\br-\br'|}.
\ee
\SB{We note in passing that within this 1l-level treatment of monovalent salt, the ion fugacities and densities are related as $\rho_{\rm b}=\Lambda_ie^{-v_{\rm b}(0)/2}$.}

In order to evaluate the integrals in Eq.~(\ref{9II}) that cannot be carried out analytically, we Taylor-expand the functions~(\ref{15}) in terms of the parameter $\gamma$ \SB{defined in Eq.~(\ref{gm})} as
\be
\label{21}
h_\pm(z)=\frac{\kappa}{p}\sum_{n\geq0}b_n^\mp e^{-v_n^\mp\tz},
\ee
where we introduced the expansion coefficients
\bea
\label{22}
b_0^\pm=u\pm1;\hspace{2mm}b^\pm_{n>0}=\pm2\gamma^{2n};\hspace{2mm}v_n^\pm=2n\pm u,
\eea
and transformed to the dimensionless wave vector as $k\to u=p/\kappa$. \SB{We note in passing that in the physiological salt conditions considered in our work where substantial screening yields $\gamma\approx1/(2s)\ll1$, the fast convergence of the series in Eq.~(\ref{21}) is assured.} 

Carrying out \SB{now} the integrals in  Eq.~(\ref{9II}) with the Green's function~(\ref{16}) and Eq.~(\ref{21}), after long algebra, the renormalized 1l-level self-energy follows in the form
\be
\label{23}
\beta\Delta\Omega^{(1)}_{\rm pp}(\tz_\p,\te)=\frac{\Gamma_{\rm s}\tau^2}{2\kappa^2}\zeta_{\rm pp}(\tz_\p,\te)
\ee
where we introduced the \RP{bulk electrostatic coupling strength}, see Table \ref{table:nonlin}, 
\be
\label{coup}
\Gamma_{\rm s}=q^2\kappa\ell_{\rm B}
\ee
with the ionic valency $q_\pm=q=1$. In Eq.~(\ref{23}), the dimensionless self-energy reads
\bea\label{24}
\zeta_{\rm pp}(\tz_\p,\te)&=&\int_0^{2\pi}\frac{\md\phi_k}{2\pi}\int_1^\infty\frac{\mathrm{d}u}{u^2-1}\left\{F(\tz_\p,\te)\right.\\
&&\left.\hspace{1cm}+\td\left[G_{\rm r}^2(\tz_\p,\te)+G_{\rm c}^2(\tz_\p,\te)\right]\right\},\nonumber
\eea
with the delta function~(\ref{17}) in dimensionless variables
\be\label{25}
\td=\frac{1+s\left(su-\sqrt{s^2+1}\right)\left(u-\eta\sqrt{u^2-1}\right)}{1+s\left(su+\sqrt{s^2+1}\right)\left(u+\eta\sqrt{u^2-1}\right)},
\ee
and the functions $F(\tz_\p,\te)$ and $G_{\rm r,c}(\tz_\p,\te)$ reported in Appendix~\ref{ap1}. 

The coupling parameter~(\ref{coup}) quantifies the importance of ion fluctuations in the salt solution and the resulting departure from the MF electrostatic regime. This parameter is related to the counterion coupling strength, see Table \ref{table:nonlin},  
\be
\Xi_{\rm c}=\frac{q^2\ell_{\rm B}}{\mu}
\ee
measuring the strength of the interfacial counterion correlations, with $\Gamma_{\rm s}=\Xi_{\rm c} s$ where $s$ is defined by Eq.~(\ref{s})~\cite{NetzSC}.

We calculate now the 1l correction to the polyelectrolyte-membrane interaction potential in Eq.~(\ref{pm1l}). Using in Eq.~(\ref{is}) the Fourier-transformed Green's function~(\ref{16}), the ionic 1l-level self-energy follows as
\bea
\label{26}
\delta v(\tz)&=&\Gamma_{\rm s}\int_1^{\infty}\frac{\mathrm{d}u}{u^2-1}\left\{-\mathrm{csch}^2\left(\tz+\tz_0\right)\right.\\
&&\hspace{2.4cm}\left.+\td\left[u+\mathrm{coth}\left(\tz+\tz_0\right)\right]^2e^{-2u\tz}\right\}.\nonumber
\eea
Inserting Eq.~(\ref{26}) into Eqs.~(\ref{chex}) and~(\ref{1lp2}) and carrying out the integral over $z'$, the 1l correction to the membrane potential follows as~\cite{Buyuk2012}
\be\label{1lp3}
\phi_{\rm m}^{(1)}(\tz)=\frac{\Gamma_{\rm s}}{4}\mathrm{csch}\left(\tz+\tz_0\right)\int_1^\infty\frac{\mathrm{d}u}{u^2-1}U(\tz),
\ee
with the auxiliary function
\bea\label{F}
U(\tz)&=&\frac{2+s^2}{s\sqrt{1+s^2}}-\td\left(\frac{1}{u}+2u+\frac{2+3s^2}{s\sqrt{1+s^2}}\right)\\
&&+\frac{\td}{u}e^{-2u\tz}+\left(\td\;e^{-2u\tz}-1\right)\coth\left(\tz+\tz_0\right).\nonumber
\eea
Substituting the potential correction~(\ref{1lp3}) into Eq.~(\ref{pm1l}) and evaluating the spatial integrals, after lengthy algebra, the 1l correction to polyelectrolyte-membrane charge coupling potential finally becomes
\be
\label{pm1l2}
\beta\Delta\Omega^{(1)}_{\rm pm}(\tz_\p,\te)=-\frac{\Gamma_{\rm s}\tau}{2\kappa}\int_1^\infty\frac{\mathrm{d}u}{u^2-1}\frac{R(\tz_\p,\te)}{\cos\te},
\ee
where the auxiliary function $R(\tz_\p)$ is given in Appendix~\ref{ap2}.

\subsection{Neutral Membranes: \SB{Repulsive Polarization and Salt-Induced "Image-Charge" Interactions}}
\label{neu}

We consider first the strict DH limit of neutral membranes with $\sigma_{\rm m}=0$ or $s\to\infty$ where the average membrane potential~(\ref{sup}) vanishes, i.e. $\phi_{\rm m}(z)=0$. As a result, the polyelectrolyte-membrane interaction potential components in Eq.~(\ref{pm1l}) vanish, $\beta\Omega^{(i)}_{\rm pm}(z_\p,\te)=0$. Consequently, the 1l polyelectrolyte grand potential~(\ref{gr1l}) reduces to the DH limit of the polyelectrolyte self-energy~(\ref{23}),
\be\label{potdh}
\beta\Delta\Omega_\p(z_\p,\te)=\beta\Delta\Omega^{\rm (DH)}_{\rm pp}(z_\p,\te)=\frac{\Gamma_{\rm s}\tau^2}{2\kappa^2}\zeta^{\rm (DH)}_{\rm pp}(\tz_\p,\te).
\ee
In Eq.~(\ref{potdh}), the DH limit of the dimensionless self-energy that follows from Eq.~(\ref{24}) reads
\bea
\label{dhself}
\zeta^{\rm (DH)}_{\rm pp}(\tz_\p,\te)&=&2\int_0^{2\pi}\frac{\md\phi_k}{2\pi}\int_1^\infty\mathrm{d}u\;\Delta_0e^{-2u\tz_\p}\\
&&\hspace{2mm}\times\frac{\cosh\left(u\tL\ct\right)-\cos(q\tL)}{u^2\cos^2\te+q^2},\nonumber
\eea
where we introduced the dielectric jump coefficient 
\be\label{del0}
\Delta_0=\frac{u-\eta\sqrt{u^2-1}}{u+\eta\sqrt{u^2-1}}
\ee
and the auxiliary function $q=\sqrt{u^2-1}\sin\te\cos\phi_k$.

\begin{figure}
\includegraphics[width=1.\linewidth]{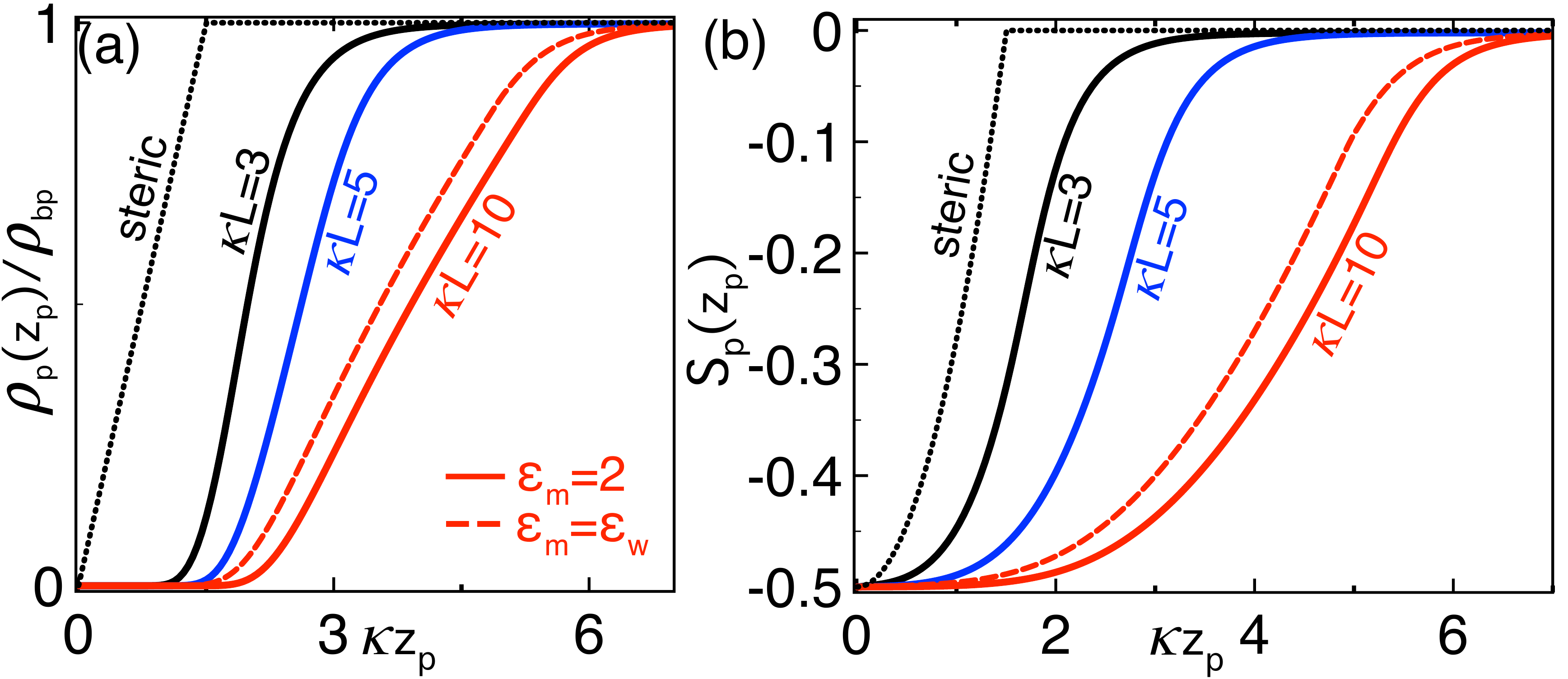}
\caption{(Color online) (a) polyelectrolyte density~(\ref{dn}) and (b) orientational order parameter~(\ref{or}) at various polyelectrolyte lengths. The neutral membrane has dielectric permittivity $\eme=2$ (solid curves) or $\eme=\ew$ (dashed red curves). The other parameters are the same as in Fig.~\ref{fig2}. The dotted black curves obtained from Eqs.~(\ref{dest}) and~(\ref{orst}) illustrate the pure steric effect associated with the rotational penalty.}
\label{fig3}
\end{figure}

Fig.~\ref{fig3}(a) displays the polyelectrolyte density~(\ref{dn}) obtained with the grand potential~(\ref{potdh}) at the biologically relevant macromolecular permittivity $\e_\m=2$ (solid curves). One notes that the electrostatic interactions between the polyelectrolyte and the neutral membrane significantly enhance the interfacial polyelectrolyte exclusion caused by the steric rotational penalty. To gain analytical insight, we focus on the far distance regime $\tz_\p\gtrsim1$ where the largest contribution to the self-energy~(\ref{dhself}) comes from the lower boundary of the integral over the variable $u$. Thus, Taylor-expanding the rational function in the second line of Eq.~(\ref{dhself}) around $u=1$, one obtains at the leading (monopolar) order
\be
\label{st1}
\zeta^{\rm (DH)}_{\rm pp}(\tz_\p,\te)\approx\kappa^2L_\p^2(\te)\int_1^\infty\mathrm{d}u\Delta_0e^{-2u\tz_\p}.
\ee
\RP{To progress further, we first consider the limit $\e_\m\ll \e_\w$, corresponding to a maximal dielectric image effect}. 
Evaluating the integral in Eq.~(\ref{st1}) in this limit, the grand potential~(\ref{potdh}) becomes
\be
\label{st2}
\beta\Delta\Omega_\p(z_\p,\te)\approx\Gamma_{\rm s}\tau^2L_\p^2(\te)\frac{e^{-2\tz_\p}}{4\tz_\p}
\ee
Eq.~(\ref{st2}) corresponds to the screened repulsive image-charge potential of an effective monopolar charge $Q_\ef(\theta)=\tau L_\p(\te)$. Hence, in this limit the polyelectrolyte depletion at the neutral dielectric membrane is driven by \RP{surface dielectric image interactions}. 

In the opposite regime of no dielectric images, {\sl i.e.}, $\e_\m=\e_\w$, the evaluation of the integral in Eq.~(\ref{st1}) yields the grand potential~(\ref{potdh}) in the form
\be\label{st3}
\beta\Delta\Omega_\p(z_\p,\te)\approx\Gamma_{\rm s}\tau^2L_\p^2(\te)\left\{\frac{(1+\tz_\p)^2}{4\tz_\p^3}-\frac{1}{2\tz_\p}\mathrm{K}_2(2\tz_\p)\right\}
\ee
where we used the modified Bessel function $K_2(x)$~\cite{math}. The polyelectrolyte energy~(\ref{st3}) corresponds to the adiabatic work required to move a point charge $Q_\ef(\theta)=\tau L_\p(\te)$ from the bulk electrolyte to the finite distance $\tz_\p$ from the neutral membrane of permittivity $\e_\m=\e_\w$~\cite{Buyuk2012}.  The corresponding repulsive ''salt-induced'' image interactions then originate solely from the charge screening deficiency of the ion-free membrane with respect to the bulk electrolyte. 

In Fig.~\ref{fig3}(a), the comparison of the solid and dashed red curves shows that the polyelectrolyte exclusion induced by this \RP{''salt-induced'' image} effect is practically as strong as the dielectric image charge exclusion. It is also noteworthy that in the strict large distance limit, $\tz_\p\gg1$, the \RP{''salt-induced'' image} potential~(\ref{st3}) tends to the dielectric image potential~(\ref{st2}), \RP{as they act in analogous ways}. Moreover, as the effective length $L_\p(\te)$ is minimized at $\te=\pi/2$ (see the inset of Fig.~\ref{fig2}(b)), Eqs.~(\ref{st2}) and~(\ref{st3}) indicate that the \RP{repulsive dielectric image and ''salt-induced'' image interactions} both tend to orient the polyelectrolyte parallel with the membrane surface. This effect is also illustrated in Fig.~\ref{fig3}(b). One sees that the interfacial region is characterized by parallel polyelectrolyte alignment, i.e. $S_\p(z_\p)<0$. Figs.~\ref{fig3}(a) and (b) also show that due to the amplification of the effective charge $Q_{\rm eff}(\theta_\p)$ and the self-energies~(\ref{st2})-(\ref{st3}), the longer the polyelectrolyte, the stronger its interfacial exclusion and parallel alignment with the membrane, i.e. $L\uparrow\rho_\p(z_\p)\downarrow S_\p(z_\p)\downarrow$. We next show that at charged membranes, these features are qualitatively modified by the interfacial counterion attraction that turns the ''salt-induced'' image interaction from repulsive to attractive.

\begin{figure*}
\includegraphics[width=1.\linewidth]{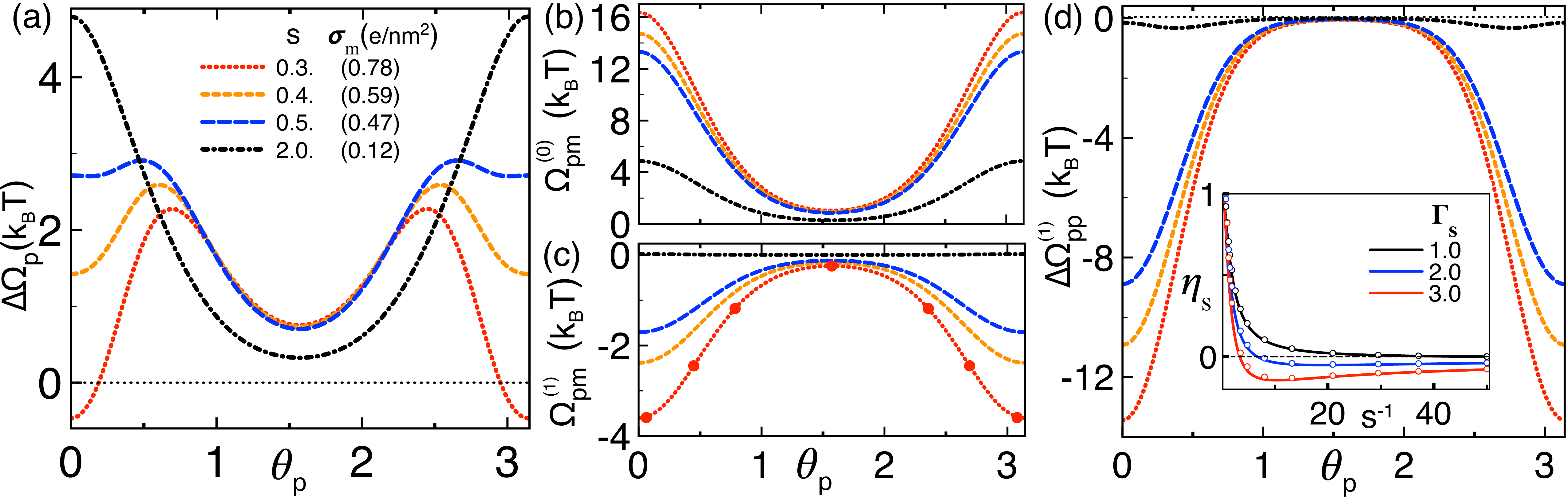}
\caption{(Color online) (a) Total polyelectrolyte grand potential~(\ref{gr1l}), (b) MF grand potential~(\ref{11}) and (c) its 1l correction in Eq.~(\ref{pm1l2}), and (d) polyelectrolyte self-energy~(\ref{23}). The dimensionless parameter $s=\kappa\mu$ for each curve is given in the legend of (a).  The red circles in (c) display the asymptotic law~(\ref{sas3}) for $s=0.3$.  The dimensionless polyelectrolyte length is $\kappa L=10$, salt density $\rho_{\rm b}=0.1$ M (coupling parameter $\Gamma_{\rm s}=0.71$), and membrane permittivity $\eme=\ew$. To eliminate the effect of the steric rotational penalty, the CM distance of the polyelectrolyte was set to the value $z_\p=0.51\;L>L/2$.  The inset in (d) illustrates the charge renormalization factor~(\ref{sas6}) (solid curves) and its analytical estimation~(\ref{sas7}) (circles) versus the dimensionless membrane charge $s^{-1}$.}
\label{fig4}
\end{figure*}

\subsection{Charged Membranes: Orientational Transition and 
\RP{Adsorption of the Polyelectrolyte}}
\label{cha}

We scrutinize here electrostatic correlations effects induced by the membrane charge on the polyelectrolyte-membrane interactions. The dielectric jump at the membrane surface is known to result in the divergence of the 1l potential correction~(\ref{1lp3})~\cite{Netz2000,Buyuk2012}. Thus, from now on, we set $\e_\m=\e_\w$, \RP{which implies no dielectric image effects and a finite ''salt-induced'' image effect}. This simplification is also motivated by recent MC studies where the surface polarization forces were observed to have a minor effect on the like-charged polyelectrolyte adsorption~\cite{Levin2016}.

\subsubsection{Intermediate Membrane Charges: Like-Charge Adsorption by ''Salt-Induced'' Image Interactions}

Figs.~\ref{fig4}(a)-(d) illustrate the total polyelectrolyte grand potential $\Delta\Omega_\p$ in Eq.~(\ref{gr1l}), the MF polyelectrolyte-membrane interaction energy $\Omega^{(0)}_{\rm pm}$ in Eq.~(\ref{11}), its 1l correction $\Omega^{(1)}_{\rm pm}$ given by Eq.~(\ref{pm1l2}), and the polyelectrolyte self-energy $\Delta\Omega^{(1)}_{\rm pp}$ in Eq.~(\ref{23}). The plots display the variation of these grand potential components with the polyelectrolyte angle $\theta_\p$ at fixed CM position $z_\p$, and for different values of the parameter $s=\kappa\mu$ ranging from the DH regime $s>1$ to the GC regime $s<1$.  \SB{The value of the polymer length $\kappa L=10$ or $L\approx9.7$ nm is comparable with the length range $10\;\mbox{nm}\lesssim L\lesssim40$ nm of the DNA molecules used in adsorption experiments~\cite{Molina2014}}.

In the DH regime $s=2$ of weak membrane charge strength or strong monovalent salt where correlation effects are negligible, i.e. $\beta\Omega^{(1)}_{\rm pm}\ll1$, $\beta\Delta\Omega^{(1)}_{\rm pp}\ll1$, and $\Delta\Omega_\p\approx\Omega^{(0)}_{\rm pm}$ (black curves), the polyelectrolyte grand potential $\Delta\Omega_\p$ is minimized by the parallel polyelectrolyte configuration $\theta_\p=\pi/2$. In Section~\ref{mf}, we showed that this originates from the repulsive polyelectrolyte-membrane charge interactions. Increasing the membrane charge or reducing the salt density, and passing to the GC regime  with $s=0.5$ and $0.4$ (blue and orange curves), the polyelectrolyte grand potential $\Delta\Omega_\p$ develops a metastable minimum at the angles $\theta_\p=\{0,\pi\}$ corresponding to the perpendicular polyelectrolyte orientation. If one moves to the stronger membrane charge regime $s=0.3$ (red curve), the perpendicular orientation becomes the stable state while the parallel orientation $\theta_\p=\pi/2$ turns to metastable. Thus, beyond a characteristic negative membrane charge strength, the anionic polyelectrolyte undergoes an orientational transition from the parallel to the perpendicular configuration. One also notes that in the same strong membrane charge regime, the grand potential in the perpendicular polyelectrolyte configuration is negative, i.e. $\Delta\Omega_\p<0$ for $\theta_\p=\{0,\pi\}$. Hence, the orientational transition of the polyelectrolyte is accompanied with its adsorption by the like-charged membrane. This is the key result of our work.

The change of the polyelectrolyte orientation upon the increment of the membrane charge strength agrees qualitatively with the conclusions of \RP{Refs.~\cite{Podgornik2009} and \cite{kim}}, where the average orientation of multipoles interacting with charged surfaces was shown to be parallel in the WC regime and perpendicular in the opposite regime of strong electrostatic coupling. In order to shed light on the physical mechanism behind the transition, we reconsider the grand potential components in Figs.~\ref{fig4}(b)-(d). These plots indicate that  the reduction of the parameter $s$ upon the rise of the membrane charge or the reduction of salt leads to two opposing effects. First, the MF grand potential component becomes more repulsive, i.e. $s\downarrow\Omega^{(0)}_{\rm pm}\uparrow$. However, the finite membrane charge also gives rise to an attractive 1l-level interaction correction $\Omega^{(1)}_{\rm pm}<0$ and polyelectrolyte self-energy $\Delta\Omega^{(1)}_{\rm pp}<0$. Figs.~\ref{fig4}(c) and (d) show that these attractive correction potentials minimized at the angles $\theta_\p=\{0,\pi\}$ favor the perpendicular polyelectrolyte configuration. Moreover, their magnitude is amplified with the membrane charge strength, i.e. $s\downarrow|\Delta\Omega^{(1)}_{\rm pp}|\uparrow|\Omega^{(1)}_{\rm pm}|\uparrow$. Consequently, beyond a critical membrane charge, the correlation-induced attractive potential components dominate the repulsive MF grand potential  $\Omega^{(0)}_{\rm pm}$. This results in the change of the polyelectrolyte orientation from parallel to perpendicular and the adsorption of the molecule by the like-charged membrane.

The attractive polyelectrolyte self-energy originates from the interfacial counterion excess that locally enhances the screening ability of the electrolyte. The stronger interfacial screening of the polyelectrolyte charges lowers the polyelectrolyte grand potential from its bulk value and thermodynamically favors the location of the molecule close to the membrane. For an analytical insight into this effect, we focus on the large distance limit $\tz_\p\gg1$ and $\tz_\p\gg\tL\ct/2$ where the largest contribution to the self-energy integral in Eq.~(\ref{24}) comes from the value of the integrand around $u=1$. Thus, we Taylor-expand the integrand of Eq.~(\ref{24}) in the neighborhood of $u=1$ and restrict ourselves to the terms of order $O\left(e^{-2\tz_\p}\right)$ in Eqs.~(\ref{a1})-(\ref{a3}). Carrying out the Fourier-integral, after lengthy algebra, the asymptotic limit of the polyelectrolyte self-energy becomes
\be
\label{sas2}
\beta\Delta\Omega_{\rm pp}^{(1)}(\tz_\p,\theta_\p)\approx-\Gamma_{\rm s} L^2\tau^2\gamma^2\left[\gamma_{\rm e}+\ln(4\tz_\p)\right]e^{-2\tz_\p},
\ee
where we used the Euler constant $\gamma_{\rm e}\approx0.57721$. The negative energy in Eq.~(\ref{sas2}) corresponds to the 1l-level attractive ''salt-induced'' image energy of a point-like ion carrying the net charge $Q=L\tau$~\cite{Buyuk2012}. Thus, for any finite membrane charge, and far enough from the substrate, the polyelectrolyte will be always subjected to a purely attractive self-energy. Then, the same enhanced screening ability of the interfacial solution leads to a negative ionic self-energy $\delta v(z)$ in Eq.~(\ref{chex}) and a positive average potential correction $\phi^{(1)}(z)>0$ in Eq.~(\ref{1lp2}). This gives rise in Eq.~(\ref{pm1l}) to a negative correction $\Omega^{(1)}_{\rm pm}<0$ to the polyelectrolyte-membrane interaction energy (see Fig.~\ref{fig4}(c)). 

\subsubsection{Strong Membrane Charges: Like-Charge Adsorption by Membrane Charge Inversion}

The like-charge adsorption effect illustrated in Fig.~\ref{fig4} is thus driven by the interfacial counterion excess. 
We now show that the like-charged polyelectrolyte-binding can be also driven by a different mechanism, namely the membrane CI. To this end, we consider the large distance regime $\tz_\p\gg1$ where Eq.~(\ref{pm1l2}) simplifies to
\be
\label{sas3}
\beta\Omega^{(1)}_{\rm pm}(z_\p,\te)\approx-\frac{\Gamma_{\rm s}}{2}{\rm I}(s)L_\p(\te)\tau e^{-\tz_\p}.
\ee
In Eq.~(\ref{sas3}), we introduced the auxiliary function
\bea
\label{sas4}
\mathrm{I}(s)&=&\int_1^\infty\frac{\mathrm{du}}{u^2-1}\left\{\frac{2+s^2}{s\sqrt{1+s^2}}-1\right.\\
&&\hspace{2cm}\left.-\td\left(\frac{1}{u}+2u+\frac{2+3s^2}{s\sqrt{1+s^2}}\right)\right\}\nonumber
\eea
and used the effective polyelectrolyte length in Eq.~(\ref{13}). The comparison of the red curve and circles in Fig.~\ref{fig4}(c) shows that Eq.~(\ref{sas3}) is accurate even close to the membrane. Using now the large distance limit of the MF grand potential~(\ref{11}), 
\be\label{sas4II}
\beta\Omega_{\rm pm}^{(0)}(z_\p,\te)\approx4\gamma L_\p(\te)\tau e^{-\tz_\p}, 
\ee
the net 1l-level polyelectrolyte-membrane charge coupling potential $\Omega_{\rm pm}=\Omega_{\rm pm}^{(0)}+\Omega_{\rm pm}^{(1)}$ takes a form similar to the DH-level MF interaction potential of Eq.~(\ref{12}),
\be
\label{sas5}
\beta\Omega_{\rm pm}(z_\p,\te)\approx\frac{2\eta_{\rm s}}{s}\tau L_\p(\te) e^{-\tz_\p}.
\ee
In Eq.~(\ref{sas5}), we introduced the membrane charge renormalization factor
\be
\label{sas6}
\eta_{\rm s}=2s\gamma\left[1-\frac{\Gamma_{\rm s}}{8}{\rm I}(s)\right]
\ee
that takes into account the effect of MF-level non-linearities and 1l-level charge correlations~\cite{Buyuk2012}.

One first notes that the 1l-level direct coupling potential~(\ref{sas5}) characterized by a longer range than the self-energy~(\ref{sas2}) dominates polyelectrolyte-membrane interactions far from the interface. Moreover, according to Eq.~(\ref{sas5}), the nature of these interactions is determined by the sign of the coefficient $\eta_{\rm s}$. This coefficient is plotted in the inset of Fig.~\ref{fig4}(d) versus the dimensionless membrane charge $s^{-1}$. For $\Gamma_{\rm s}\lesssim1$, due to the enhancement of MF-level non-linearities, the increment of the membrane charge reduces the purely positive renormalization factor from $\eta_{\rm s}=1$ to $0$. At larger coupling parameters  $\Gamma_{\rm s}\gtrsim1$, beyond a characteristic membrane charge $s_*^{-1}$, $\eta_{\rm s}$ turns from positive to negative. This corresponds to the membrane CI phenomenon. Consequently, the potential~(\ref{sas5}) characterizing polyelectrolyte-membrane interactions far from the interface switches from repulsive to attractive, indicating the polyelectrolyte attraction by the like-charged substrate.

To identify the CI point, we evaluate the integral~(\ref{sas4}) in the GC regime $s\ll1$ to obtain $I(s)\approx-2\ln(s)$ and
\be
\label{sas7}
\eta_{\rm s}\approx2s\gamma\left[1+\frac{\Gamma_{\rm s}}{4}\ln(s)\right].
\ee
Eq.~(\ref{sas7}) reported in the inset of Fig.~\ref{fig4}(d) by circles can accurately reproduce the trend of the renormalization coefficient~(\ref{sas6}). According to Eq.~(\ref{sas7}), CI occurs at the dimensionless inverse membrane charge 
\be\label{sas8}
s_*=e^{-4/\Gamma_{\rm s}} .
\ee
In agreement with the inset of Fig.~\ref{fig4}(d), Eq.~(\ref{sas8}) indicates the decrease of this critical membrane charge with the coupling parameter, i.e. $\Gamma_{\rm s}\uparrow s^{-1}_*\downarrow$. 

At this point, we emphasize that in Fig.~\ref{fig4}, the like charge adsorption at $s=0.3\gg s_*\approx3.7\times10^{-3}$ takes place without the occurrence of the CI. This shows the absence of one-to-one mapping between the membrane CI and the like-charge polyelectrolyte-membrane complexation driven by the ''salt-induced'' image interaction excess; in agreement with the observation of recent Monte-Carlo (MC) simulations~\cite{Levin2016}, the like-charge polyelectrolyte binding may occur at membrane charge strengths well below the threshold~(\ref{sas8}) required for the onset of the CI. To summarize, at moderate membrane charges $s>s_*$, the like-charge polyelectrolyte binding can occur exclusively as a result of \SB{the salt-induced "image-charge" effect enhanced} by the dense cations in the close vicinity of the membrane. In the strong membrane charge regime $s<s_*$, the membrane CI will act as an additional mechanism capable of inducing the polyelectrolyte adsorption over an extended distance from the membrane surface.

\begin{figure}
\includegraphics[width=.9\linewidth]{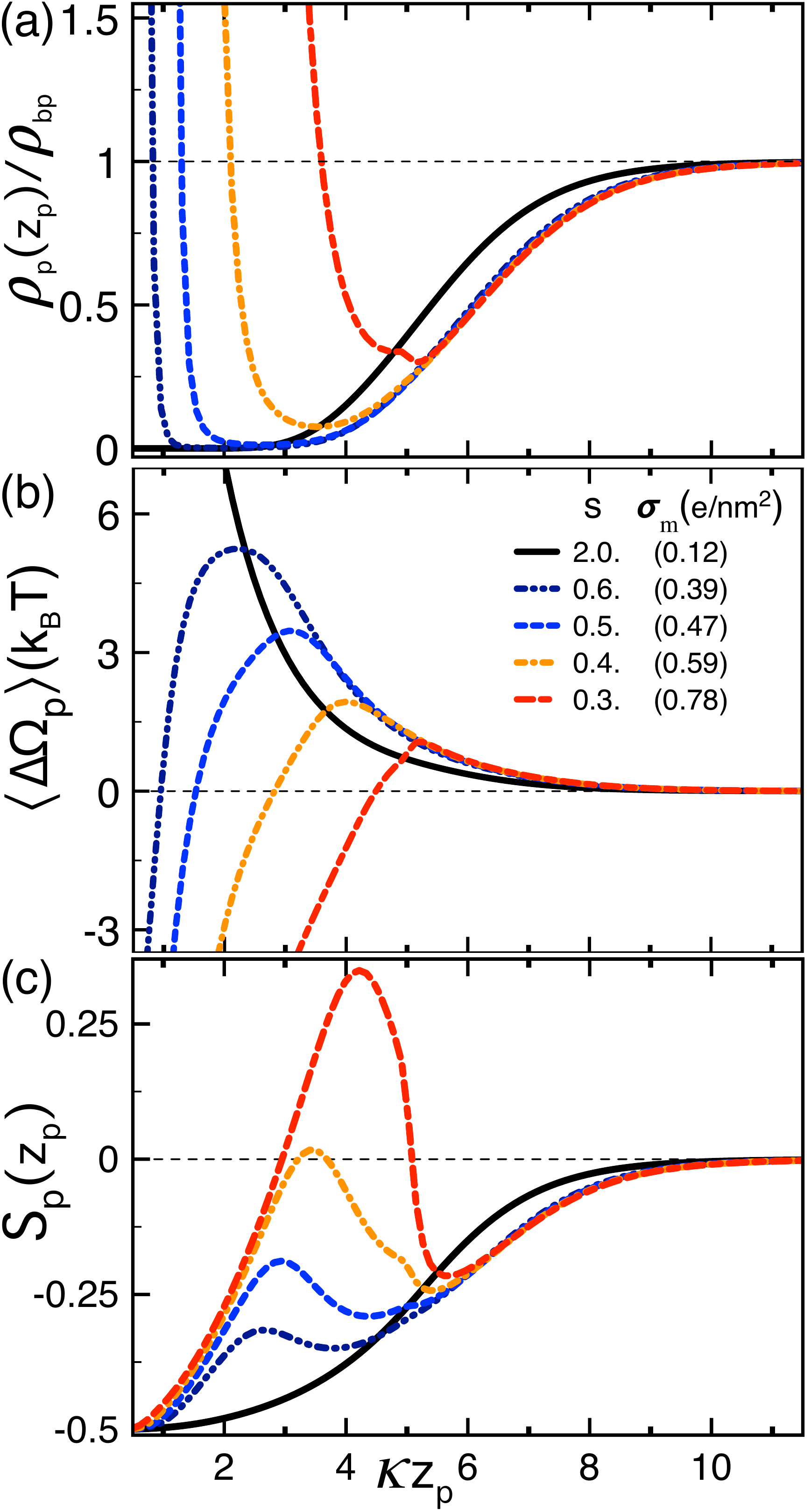}
\caption{(Color online)  (a) polyelectrolyte density~(\ref{dn}), (b) polyelectrolyte grand potential~(\ref{gr1l}) averaged over polyelectrolyte rotations, and (c) orientational order parameter~(\ref{or}). The dimensionless parameter $s$ and the corresponding membrane charge $\sigma_{\rm m}$ for each curve is given in the legend of (b).  The dimensionless polyelectrolyte length is $\kappa L=10$ and the salt density $\rho_{\rm b}=0.1$ M.}
\label{fig5}
\end{figure}

\subsubsection{Interfacial Polyelectrolyte Configuration at the Transition}

We investigate here the interfacial polyelectrolyte configuration in the polyelectrolyte adsorption regime. Figs.~\ref{fig5}(a)-(b) display the polyelectrolyte density profile, and the grand potential averaged over polyelectrolyte rotations according to Eq.~(\ref{defi}). In the DH regime $s=2$ (black curves), the MF-level polyelectrolyte-membrane repulsion leads to a pure interfacial polyelectrolyte depletion $\rho_\p(z_\p)<\rho_{\rm bp}$. Passing to the GC regime of stronger membrane charges $s\lesssim0.6$, the polyelectrolyte grand potential keeps its repulsive branch far from the interface but the correlation ''salt-induced'' image interactions give rise to an additional attractive branch in the close vicinity of the membrane surface. This leads to a piecewise polyelectrolyte configuration characterized by polyelectrolyte adsorption $\rho_\p(z_\p)>\rho_{\rm bp}$ over the interfacial layer of width $d$, which is followed by a polyelectrolyte depletion layer $\rho_\p(z_\p)<\rho_{\rm bp}$ at $z_\p>d$. Figs.~\ref{fig5}(a)-(b) also show that the stronger the membrane charge, the more attractive the average grand potential, and the larger the adsorbed polyelectrolyte layer, i.e. $s\downarrow\lan\Delta\Omega_p(z_\p)\ran\downarrow d\uparrow$. This result agrees with the experiments of Ref.~\cite{Molina2014} where the density of dsDNA molecules adsorbed onto anionic lipid monolayers was found to be higher in the dipalmitoylphosphatidyslerine  rich regions of the substrate characterized by a stronger surface charge.

Fig.~\ref{fig5}(c) displays the effect of charge correlations on the polyelectrolyte orientation profile. In the weak membrane charge regime $s=2$, the system is characterized by the MF behavior of parallel polyelectrolyte alignment $S_\p(z_\p)<0$ along the membrane surface. Rising the membrane charge into the GC regime $s\lesssim1$ (navy and blue curves), the onset of like-charge attraction very close to the interface gives rise to the peak of the order parameter $S_\p(z_\p)$. This indicates the tendency of the polyelectrolyte to orient itself perpendicular to the membrane. In the stronger membrane charge regime $s\lesssim0.4$ where attractive \SB{salt-induced "image-charge"} forces become comparable with the MF repulsion, the orientational order profile exhibits an oscillatory behavior. Namely, away from the surface where the grand potential is repulsive, the order parameter indicates parallel polyelectrolyte alignment $S_\p(z_\p)<0$. As one approaches the interface and gets into the layer where the grand potential has an attractive branch, the order parameter sharply rises and reaches the regime $S_\p(z_\p)>0$ indicating the transition of the polyelectrolyte orientation from parallel to perpendicular. Then, in the immediate vicinity of the membrane surface $z\lesssim L/2$ where the steric rotational penalty comes into play, the order parameter drops again below the limit $S_\p(z_\p)=0$  and the polyelectrolyte orientation switches from perpendicular back to parallel. 

\begin{figure}
\includegraphics[width=.9\linewidth]{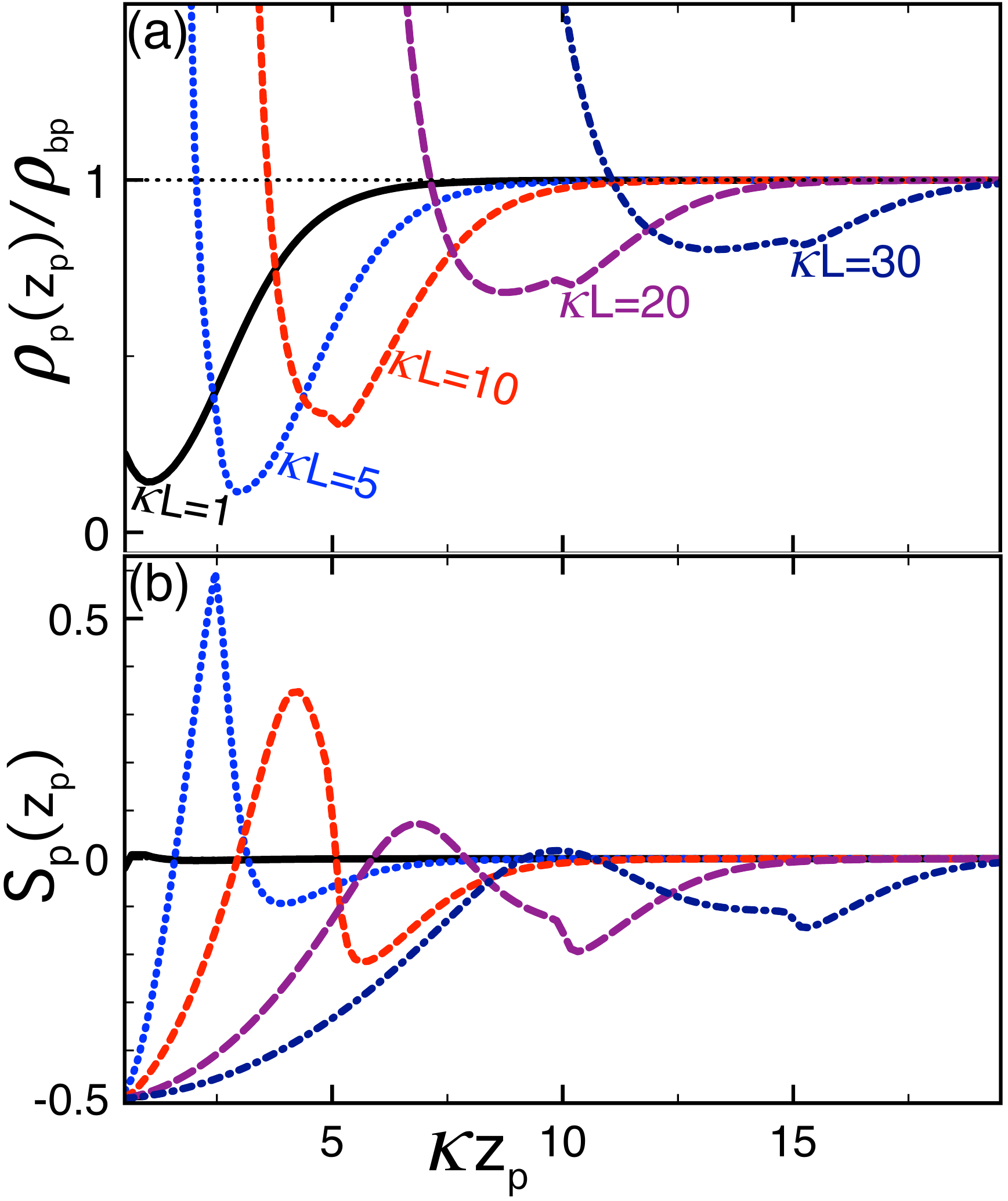}
\caption{(Color online) (a) polyelectrolyte density~(\ref{dn}) and (b) orientational order parameter~(\ref{or}) at the dimensionless parameter $s=0.3$ and for various polyelectrolyte lengths indicated in (a).}
\label{fig6}
\end{figure}

\RP{The extension of the polyelectrolyte length, implying also an increase of the polyelectrolyte charge}, amplifies both the MF-level like-charge polyelectrolyte-membrane repulsion and the opposing ''salt-induced' image interaction attraction. In order to understand the net effect of the polyelectrolyte size, in Figs.~\ref{fig6}(a) and (b), we reported the polyelectrolyte density and orientational order parameter at various polyelectrolyte lengths. First of all, Fig.~\ref{fig6}(a) shows that polyelectrolyte adsorption occurs only if the polyelectrolyte length is above a minimum threshold, i.e.  $\rho_\p(z_\p)>\rho_{\rm pb}$ if $L\gtrsim\kappa^{-1}$. Then, one notes that the longer the polyelectrolyte, the wider the adsorption layer, and the larger the adsorbed polyelectrolyte density, i.e. $L\uparrow d\uparrow\rho_\p(z_\p)\uparrow$. 

Thus, the overall effect of the polyelectrolyte length extension is the monotonical enhancement of the correlation-induced attraction. However, Fig.~\ref{fig6}(b) shows that the orientational order depends on the polyelectrolyte length in a non-monotonic fashion. Namely, increasing the length of the molecule from $L=\kappa^{-1}$ to $L=L_{\rm c}=5\;\kappa^{-1}$, the amplification of attractive \SB{salt-induced image-charge} forces rises sharply the order parameter ($L\uparrow S_\p(z_\p)\uparrow$) and turns the interfacial polyelectrolyte orientation to perpendicular. This trend is however reversed beyond the characteristic length $L_{\rm c}$; for $L>L_{\rm c}$, the interfacial layer $z_\p<L/2$ associated with the steric rotational penalty covers the attractive grand potential layer responsible for the perpendicular polyelectrolyte alignment. As a result, the extension of the polyelectrolyte length beyond $L\approx L_{\rm c}$ drops the peak of the order parameter ($L\uparrow S_\p(z_\p)\downarrow$) and decreases the tendency of the polyelectrolyte to orient itself perpendicular to the membrane. To summarize, the like-charge polyelectrolyte binding is accompanied with the orientational transition only up to a critical polyelectrolyte length ($L\approx20\;\kappa^{-1}$ in Fig.~\ref{fig6}(b)). Due to the steric penalty, the adsorption of longer polyelectrolytes occurs without the orientational transition of the molecule.

\SB{We finally note that our analysis of polymer-membrane interactions in the salt solution was based on a perturbative treatment of the polymer charge. As this composite charged system of considerable complexity includes several characteristic lengths, a simple dimensional analysis that would enable the quantitatively reliable determination of the validity regime of this perturbative approximation is not possible. This indicates that an accurate identification of the validity regime of the test charge approach should be done in a future work by extensive comparisons with simulations and/or by a test charge theory of higher perturbative level. Such an extension is of course beyond the scope of the present work.}

\section{1l correlations in mono- and divalent counterion-only liquids}
\label{counon}

With the aim to gain further analytical insight into the correlation effects observed in Section~\ref{corsalt} and to understand the role of the cation valency on the adsorption transition, we investigate here polyelectrolyte-membrane interactions in mono- and divalent counterion-only liquids.

\subsection{Derivation of the Electrostatic Ion Potentials}
\label{pi}

For the computation of the polyelectrolyte potentials in the counterion-only liquid, we briefly review here the derivation of the ionic potentials $\phi^{(i)}_\m(z)$ and $v(\br,\br'$) calculated in Ref.~\cite{Netz2000}. We set the membrane permittivity to $\e_{\rm m}=\e_{\rm w}$. First, the solution to the PB Eq.~(\ref{4}) is
\be\label{co1}
\phi^{(0)}_\m(\bz)=\frac{2}{q}\ln\left(1+\bz\right)
\ee
where we introduced the dimensionless distance $\bz=z/\mu$. Hence, the counterion density satisfying the electroneutrality condition $q\int_0^\infty\mathrm{d}zn(z)=\sigma_\m$ becomes
\be
\label{co2}
n(\bz)=\frac{2\pi\ell_{\rm B}\sigma_\m^2}{\left(1+\bz\right)^2}.
\ee

Substituting Eq.~(\ref{co2}) into Eq.~(\ref{ep}), the differential equation~(\ref{eh}) takes the form
\be
\label{co3}
h''_{\pm}(z)-\left\{k^2+\frac{2}{\left(\mu+z\right)^2}\right\}h_{\pm}(z)=0.
\ee
The solution to Eq.~(\ref{co3}) reads~\cite{Netz2000}
\be\label{co4}
h_\pm(z)=e^{\pm kz}\left(k\mp\frac{1}{z+\mu}\right).
\ee
Injecting Eq.~(\ref{co4}) into the general solution in Eq.~(\ref{8II}), the Fourier-transformed Green's function becomes
\be\label{co5}
\tv(z,z')=\frac{2\pi\ell_{\rm B}}{k^3}\left[h_+(z_<)+\Delta h_-(z_<)\right]h_-(z_>),
\ee
with the delta function $\Delta=\left(1+2\rk+2\rk^2\right)^{-1}$ and the dimensionless wave vector $\rk=\mu k$. Using Eq.~(\ref{co5}) in Eq.~(\ref{ogpp}), the 1l ionic self-energy takes the integral form
\bea\label{co6}
\delta v(\bz)&=&\frac{\ell_{\rm B}}{\mu}\int_0^\infty\frac{\mathrm{d}\rk}{\rk^2}\left\{-\frac{1}{\left(1+\bz\right)^2}\right.\\
&&\hspace{2cm}\left.+\Delta\left(\rk+\frac{1}{1+\bz}\right)^2e^{-2\rk\bz}\right\}.\nonumber
\eea
Finally, substituting Eqs.~(\ref{co2}), (\ref{co5}), and~(\ref{co6}) into Eq.~(\ref{1lp2}), and carrying-out the spatial integral, the 1l correction to the average potential follows as
\bea\label{co7}
\phi^{(1)}_\m(\bz)&=&\frac{q\ell_{\rm B}}{4\mu\left(1+\bz\right)^2}\int_0^\infty\frac{\mathrm{d}\rk}{\rk^2}\left\{-2\bz\left[\Delta(1+\rk)-1\right]\right.\\
&&\hspace{2.7cm}\left.+1+\Delta\left(e^{-2\rk\bz}-2\rk-2\right)\right\}.\nonumber
\eea

\subsection{Polyelectrolyte Adsorption in the Counterion Liquid}

In the counterion-only liquid, due to the long range of the unscreened polyelectrolyte-membrane interactions, the total interaction potential in Eq.~(\ref{gr1l}) is weakly affected by the orientational configuration of the molecule. The corresponding results presented in Appendix~\ref{pi} will not be reported here. Based on this observation, we simplify the following analysis by restricting ourselves to the parallel polyelectrolyte orientation and set $\te=\pi/2$.

\begin{figure}
\includegraphics[width=1\linewidth]{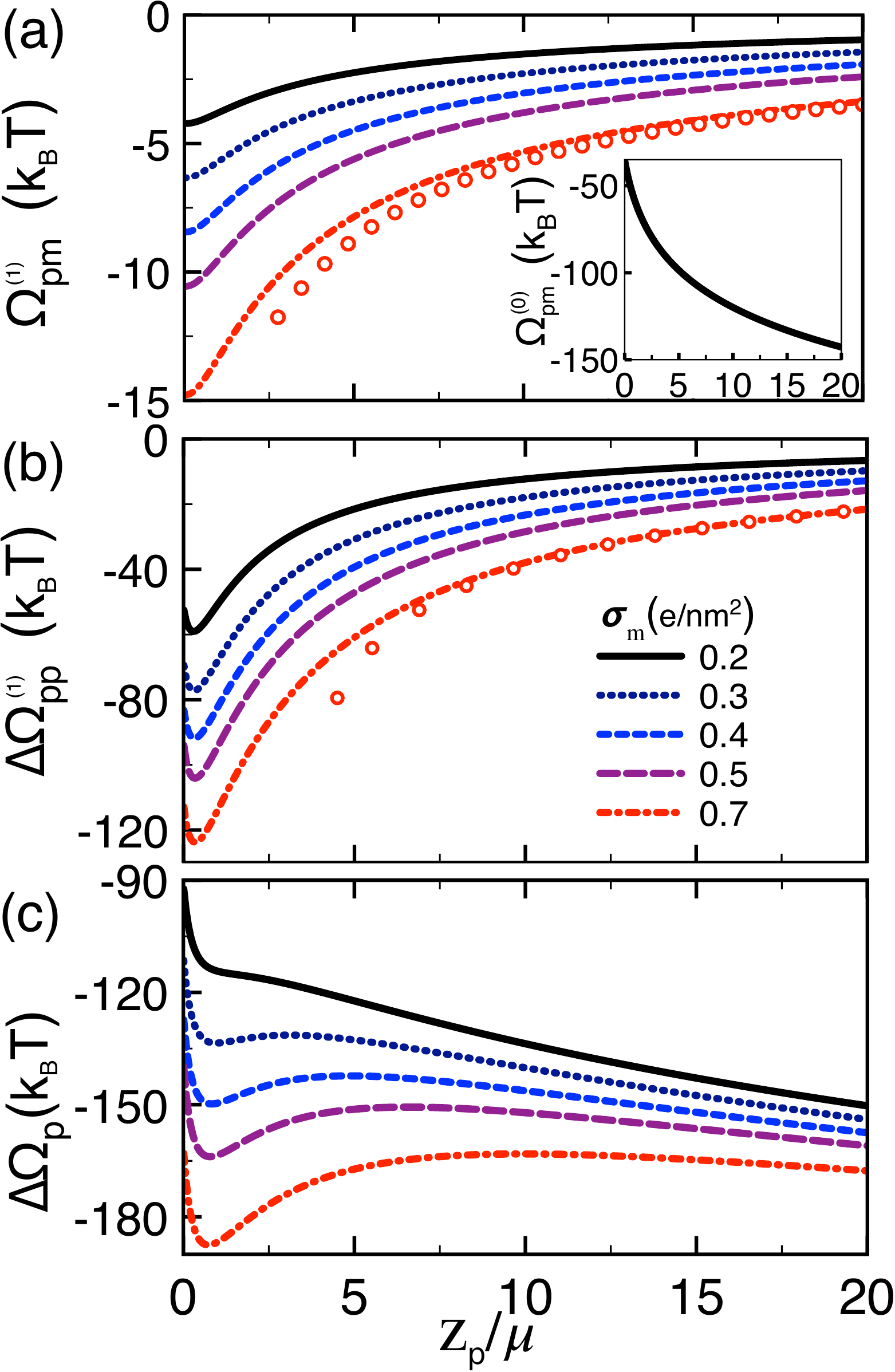}
\caption{(Color online)  (a) MF polyelectrolyte-membrane interaction potential~(\ref{couPMF}) (inset), its 1l correction~(\ref{couP1l}) (curves in the main plot), and the asymptotic limit~(\ref{co20}) (circles). (b) polyelectrolyte self-energy~(\ref{co23}) (curves) and its large distance limit~(\ref{co26}) (circles). (c) Total grand potential profile~(\ref{gr1l}). The liquid is monovalent (q=1). The polyelectrolyte angle is $\te=\pi/2$ and length $L=3$ nm. The membrane charge densities for each curve is given in the legend of (b).}
\label{fig7}
\end{figure}

The MF-level polyelectrolyte-membrane interaction energy follows by inserting the MF potential~(\ref{co1}) into Eq.~(\ref{pm1l}) and carrying out the integral. This yields
\be
\label{couPMF}
\beta\Omega^{(0)}_{\rm pm}(\bzp,\te)=-\frac{2Q}{q}\left[1+\ln\left(1+\bzp\right)\right],
\ee
with the polyelectrolyte charge $Q=L\tau$ and the dimensionless polyelectrolyte distance $\bzp=z_\p/\mu$. To compute the 1l correction to the MF energy~(\ref{couPMF}), we substitute into Eq.~(\ref{pm1l}) the average potential correction~(\ref{co7}). One finds
\bea
\label{couP1l}
\beta\Omega^{(1)}_{\rm pm}(\bzp,\te)&=&\frac{\Xi_{\rm c} Q}{8q\left(1+\bzp\right)^2}\left\{4\bzp-4\pi e^{\bzp}\sin(\bzp)\right.\\
&&\hspace{1.cm}-\left[4\gamma_{\rm e}+\pi+\ln\left(4\bzp^4\right)\right](1+\bzp)\nonumber\\
&&\hspace{1.cm}\left.+4\;\mathrm{Re}\left[e^{(1+i)\bzp}\mathrm{Ei}\left[-(1+i)\bzp\right]\right]\right\}\nonumber
\eea
where we used the exponential integral function $\mathrm{Ei}(x)$~\cite{math}. Fig.~\ref{fig7}(a) shows for monovalent counterions the landscape of the repulsive MF potential~(\ref{couPMF}) driving the polyelectrolyte away from the membrane (inset), and its 1l correction~(\ref{couP1l}) of uphill trend attracting the polyelectrolyte towards the substrate (main plot). In the strict large distance limit $\bzp\gg1$, Eq.~(\ref{couP1l}) tends to the limiting law 
\be\label{co20}
\beta\Omega^{(1)}_{\rm pm}(\bzp,\te)\approx-\frac{\Xi_{\rm c} Q}{8q\bzp}\left[-4+4\gamma_{\rm e}+\pi+\ln\left(4\bzp^4\right)\right]
\ee
displayed in Fig.~\ref{fig7}(a) by circles. Eq.~(\ref{co20}) shows that the correction potential $\Omega^{(1)}_{\rm pm}$ is purely attractive and it decays inversely with the polyelectrolyte distance. 

In order to derive the polyelectrolyte self-energy, we insert the Green's function in Eq.~(\ref{co5}) into Eq.~(\ref{9II}) to obtain
\bea\label{co23}
\beta\Delta\Omega^{(1)}_{\rm pp}(\bzp)&=&\frac{\Xi_{\rm c} Q^2}{2q^2}\int_0^\infty\frac{\mathrm{d}\rk}{\rk^2} P(\rk\bL)\\
&&\hspace{-4mm}\times\left\{-\frac{1}{\left(1+\bz\right)^2}+\Delta\left(\rk+\frac{1}{1+\bz}\right)^2e^{-2\rk\bz}\right\},\nonumber
\eea
with the dimensionless polyelectrolyte length $\bL=L/\mu$ and the polyelectrolyte structure factor
\be
\label{co24}
P(x)=\left[\pi{\rm H}_0(x)-\frac{2}{x}\right]{\rm J}_1(x)+\left[2-\pi{\rm H}_1(x)\right]{\rm J}_0(x)
\ee
where we used the Struve function ${\rm H}_n(x)$ and the Bessel function $J_n(x)$~\cite{math}. In the short polyelectrolyte regime $\bL\ll1$ where $P(\rk\bL)\to1$, Eq.~(\ref{co23}) tends to the self-energy~(\ref{co6}) of a point charge $Q$, i.e. $\Delta\Omega_{\rm pp}(\bzp)\to Q^2\delta v(\bzp)/2$. Then, at large distances $z_{\rm p}+\mu\gg L|\ct|/2$, Eq.~(\ref{co23}) takes the asymptotic form 
\bea
\label{co26}
\beta\Delta\Omega_{\rm pp}^{(1)}(\bzp,\te)&\approx&\frac{\Xi_{\rm c} Q^2}{q^2}\left\{-\frac{3}{4\bzp}+\frac{2\bL+9}{12\bzp^2}\right.\\
&&\left.\hspace{1.2cm}-\frac{5\bL^2+64\bL+144}{192\bzp^3}\right\}.\nonumber
\eea
Eqs.~(\ref{co23}) and~(\ref{co26}) displayed in Fig.~\ref{fig7}(b) indicate that due to the locally enhanced screening by the interfacial cations, the polyelectrolyte self-energy is attractive and it decays algebraically with the polyelectrolyte distance $z_{\rm p}$.  One also notes that its magnitude is an order of magnitude higher than the potential correction $\Omega^{(1)}_{\rm pm}$ in Fig.~\ref{fig7}(a). Thus, in counterion liquids, the self-energy brings the main attractive contribution to polyelectrolyte-membrane interactions.

In Fig.~\ref{fig7}(c), one notes that in the weak membrane charge regime $\sigma_{\rm m}\lesssim0.2$ $e/{\rm nm}^2$ governed by the MF interaction potential~(\ref{couPMF}), the total 1l grand potential $\Delta\Omega_{\rm p}$ is repulsive (black curve). Then, Figs.~\ref{fig7}(a) and (b) show that the rise of the membrane charge enhances the interfacial counterion density and amplifies the attractive 1l correction potentials, i.e. $\sigma_{\rm m}\uparrow\Omega^{(1)}_{\rm pm}\downarrow\Delta\Omega_{\rm pp}^{(1)}\downarrow$. As a result, close to the membrane, the total grand potential develops an attractive well whose depth increases with the membrane charge strength, i.e. $\sigma_{\rm m}\uparrow\Delta\Omega_{\rm p}\downarrow$. This is the signature of the like-charge polyelectrolyte adsorption. However, far enough from the membrane, the repulsive MF interaction potential~(\ref{couPMF}) growing logarithmically with the distance dominates the attractive potential components~(\ref{co20}) and~(\ref{co26}) decaying algebraically. This leads to the downhill landscape of the grand potential $\Delta\Omega_{\rm p}$ at $\bz_{\rm p}\gg1$. Hence, in counterion-only solutions, the polyelectrolyte located at sufficiently large distances will be always repelled by the membrane. This is due to the non-occurrence of CI in counterion-only liquids and the absence of the CI-driven long-ranged like-charge attraction mechanism observed in Sec.~\ref{cha} with finite salt. Thus, in counterion liquids, the like-charge adsorption can take place solely due to the short-ranged enhanced ''salt-induced'' image interactions due to the high cation density close to the membrane.

\SB{It should be finally noted that due to the perturbative treatment of the polymer charge, the \textit{Manning-Osawa condensation} is not taken into account by our test charge formalism. The consideration of this non-linear electrostatic effect originating from strong counterion condensation requires  the non-perturbative treatment of the polymer charge. This extension discussed in Conclusions lies beyond the scope of the present work.}

\subsection{Effect of the Polyelectrolyte Length and Charge, and Ion Valency}

\begin{figure}
\includegraphics[width=1\linewidth]{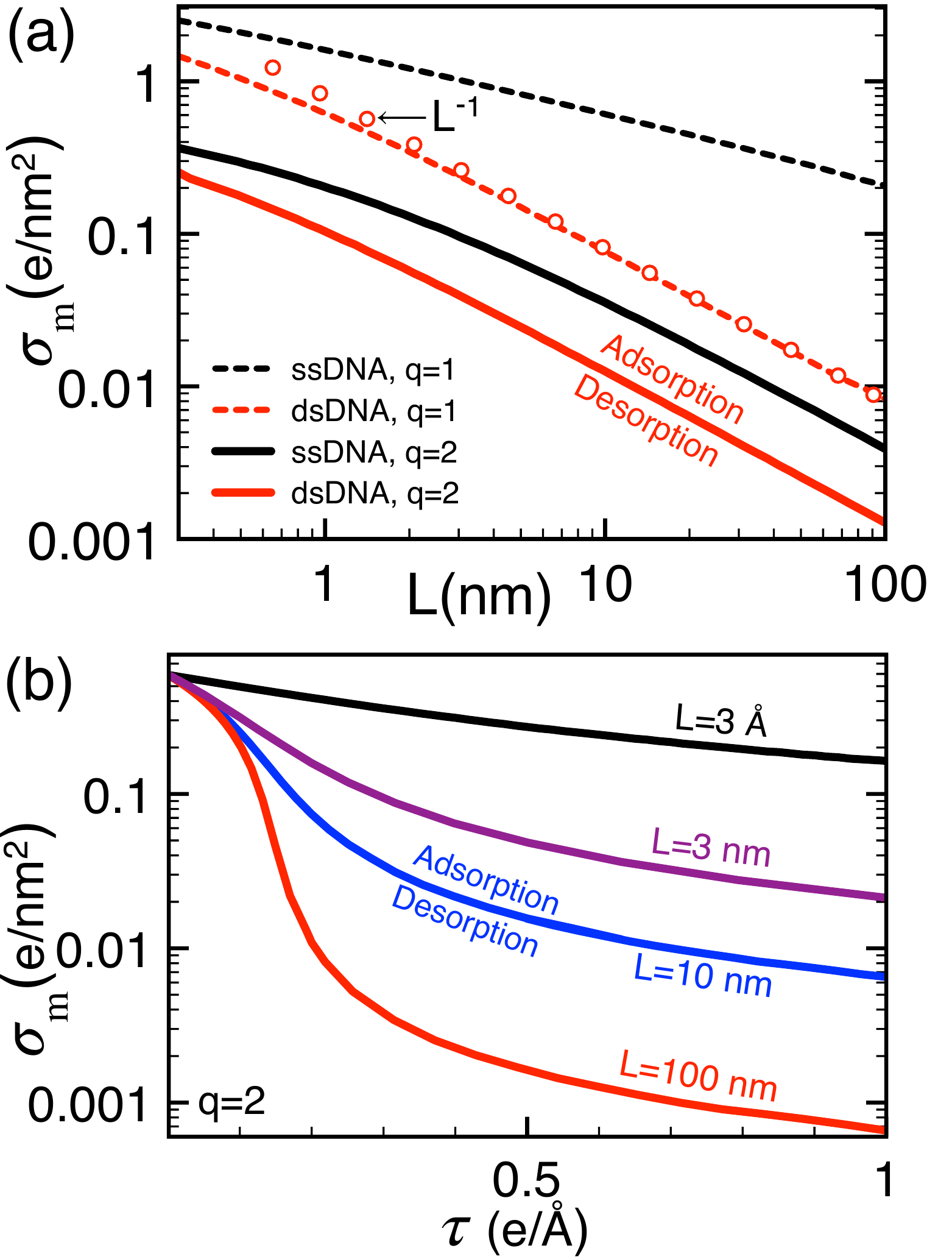}
\caption{(Color online) The critical membrane charge $\sigma^*_{\rm m}$ for the onset of the like-charge polyelectrolyte adsorption versus (a) the length $L$ and (b) charge density $\tau$ of the molecule. The curves separating the attractive phase (area above the curves) and repulsive phase (below the curves) are plotted for monovalent (dashed curves) and divalent counterions (solid curves). The curves in (a) are for the ssDNA charge density $\tau=1/(3.4\;\mbox{{\AA}})$ (black) and the twice higher dsDNA charge density (red). }
\label{fig8}
\end{figure}

Fig.~\ref{fig8}(a) displays the critical membrane charge $\sigma_{\rm m}^*$ for the onset of the like-charge polyelectrolyte-membrane attraction versus the polyelectrolyte length $L$. The result is computed for ssDNA (black curves) and dsDNA of twice higher charge (red curves) in monovalent (dashed curves) and divalent counterions (solid curves). First, the phase diagram shows that the longer the polyelectrolyte, the lower the critical membrane charge, i.e. $L\uparrow\sigma_{\rm m}^*\downarrow$. Thus, the extension of the molecule favors its adsorption. This peculiarity can be explained by the competition between the repulsive MF interaction potential~(\ref{couPMF}) linear in $L$ and the attractive self-energy~(\ref{co26}) whose leading order term grows quadratically with $L$. The ratio of these potentials scaling as $\Delta\Omega^{(1)}_{\rm pp}/\Omega^{(0)}_{\rm pm}\propto\sigma_{\rm m}L$ implies that in order to keep intact the strength of the attractive interactions, any reduction of the charge $\sigma_{\rm m}$ should be compensated by the extension of the length $L$ by the same factor. Hence, the critical membrane charge should behave with the length $L$ as  $\sigma_{\rm m}^*\sim L^{-1}$. Fig.~\ref{fig8}(a) shows that this scaling law characterizes accurately the long polyelectrolyte regime of all critical lines, except the case of ssDNA in monovalent liquids where the transition regime to the inverse linear scaling extends beyond the range of the figure.

In Fig.~\ref{fig8}(a), we also illustrate the role played by the counterion valency in the polyelectrolyte adsorption. The comparison of the solid and dashed curves shows that in the divalent counterion liquid, the like-charge adsorption of ssDNA and dsDNA molecules both occur at membranes of an order of magnitude lower charge density than with monovalent counterions, i.e. $q\uparrow\sigma_{\rm m}^*\downarrow$. This feature stems from the enhancement of the screening ability of the interfacial counterions with the increase of their valency. Such a tendency has been indeed observed in adsorption experiments~\cite{Molina2014,Qiu2015,Tiraferri2015,Fries2017} and simulations~\cite{Levin2016} where the multivalency of counterions was found to facilitate the complexation of DNA molecules with anionic lipid monolayers.

Finally, we investigate the overall effect of the polyelectrolyte charge strength on the adsorption transition. In Fig.~\ref{fig8}(a), the comparison of the black and red curves indicates that at fixed polyelectrolyte length, dsDNA molecules are adsorbed at significantly lower membrane charges than ssDNA molecules. This trend is also illustrated in Fig.~\ref{fig8}(b). One notes that the critical membrane charge drops monotonically with the increment of the polyelectrolyte charge ($\tau\uparrow\sigma_{\rm m}^*\downarrow$), and the effect is strongly amplified by the polyelectrolyte length $L$. This behavior is again due to the competition between the repulsive MF potential~(\ref{couPMF}) and the attractive self-energy~(\ref{co23}); the increment of the polyelectrolyte charge brings a stronger contribution to the self energy quadratic in $\tau$. We however note that the validity of this conclusion is limited by our treatment of the polyelectrolyte charge at the quadratic order. The extension of the present test-charge approach beyond the quadratic approximation or numerical simulations will be needed for the confirmation of this prediction.

\section{Conclusions}

The optimization of modern biosensing and genetic engineering approaches requires an accurate insight into the behavior of biopolyelectrolytes interacting with charged macromolecules. In this work, we characterized the interaction of anionic polyelectrolytes with like-charged membranes in the presence of mobile ions. From gene delivery techniques to nanopore-based  sequencing strategies, our model is relevant to various biotechnological methods involving polyelectrolyte-membrane complexes.

Our characterization of polyelectrolyte-membrane interactions was based on a generalized test charge formalism. This approach consists of expanding the electrostatic partition function of the system at the quadratic order in the charge density of the rotating stiff polyelectrolyte. Within this systematic perturbative expansion, we derived the polyelectrolyte grand potential dressed by the exact electrostatic ion-membrane interactions. In order to put this grand potential in an analytically tractable form, we formulated the polyelectrolyte-membrane interactions within the 1l theory of inhomogeneous electrolytes. In terms of this 1l-level polyelectrolyte grand potential, we investigated the effect of charge correlations on the configuration of the anionic polyelectrolyte interacting with a like-charged membrane. 

We found that polyelectrolyte-membrane interactions are governed by the direct coupling of the polyelectrolyte charges with the cation-dressed membrane charges, and the ''salt-induced'' image interactions in the non-uniformly partitioned electrolyte solution. In weakly charged membranes,  
the repulsive polyelectrolyte-membrane interactions and ''salt-induced'' image interactions  lead to the polyelectrolyte exclusion from the interfacial region and the parallel orientation of the molecule with the membrane surface. At intermediate membrane charges, 
the interfacial screening excess originating from the cation attraction to the surface turns the ''salt-induced'' image interactions and the net polyelectrolyte grand potential from repulsive to attractive. As a result, the polyelectrolyte undergoes an orientational transition from parallel to perpendicular configuration, which is accompanied with the like-charge adsorption of the molecule by the membrane. Finally, in the \RP{stronger membrane charge regime, but still at intermediate coupling}, the emerging membrane CI acts as an additional mechanism capable of triggering the like-charged polyelectrolyte adsorption over an extended distance from the membrane. 

In agreement with adsorption experiments, we showed that the like-charged polyelectrolyte adsorption effect is amplified by both the membrane charge strength and the ion multivalency. Our investigation revealed that the extension of the polyelectrolyte length also favors the binding of the molecule onto the similarly charged substrate. However, due to the steric penalty, the adsorption of the polyelectrolyte is accompanied by its orientational transition only up to a critical polyelectrolyte length corresponding roughly to the range of the interfacial \SB{salt-induced image-charge} forces. 

In this work, we focused exclusively on the case of mono- and divalent electrolytes. Thus, our electrostatic formalism was based on the 1l theory of inhomogeneous solutions able to cover the corresponding weak to intermediate electrostatic coupling regime.  We emphasize that since the validity of the generalized test charge \SB{approach} does not depend on the strength of the electrolyte-membrane coupling, the theory can be readily applied to understand polyelectrolyte-membrane interactions \SB{in solutions including tri- and tetravalent cations. The importance of this extension stems from the fact that adsorption experiments often involve the mixture of high valency counterions with monovalent salt. It should be however noted that the strong coupling interactions arising from tri- and tetravalent ions lie beyond the reach of the present 1l theory. It is indeed known that in the electrostatic strong-coupling regime, the loop expansion of the liquid grand potential is not convergent~\cite{NetzSC}. Therefore, the consideration of high valency counterions will require the use of a strong coupling approach such as the virial expansion of the liquid grand potential in terms of the multivalent charge fugacity~\cite{Podgornik2010}.} Our study of the system in this strong coupling regime will be presented in an upcoming work. We finally note that the test charge theory is based on a perturbative treatment of the polyelectrolyte charge at the quadratic order. We plan \SB{to identify quantitatively in a future work the validity regime of the corresponding approximation by systematic comparisons} with MC simulations and an improved test charge theory of higher order perturbative level.

\smallskip
\appendix

\begin{widetext}

\section{Auxiliary Functions $F(\tz_\p,\te)$ and $G_{\rm r,c}(\tz_\p,\te)$ of the Polyelectrolyte Self-energy~(\ref{24})}
\label{ap1}

We report here the auxiliary functions $F(\tz_\p,\te)$ and $G_{\rm r,c}(\tz_\p,\te)$ of the polyelectrolyte self-energy~(\ref{24}).
\bea\label{a1}
G_{\rm r}(\tz_\p,\te)&=&2\sum_{n=0}^\infty\frac{b_n^+}{t{_n^+}^2+\alpha^2}\left\{t_n^+\sinh\left(\frac{t_n^+\tL}{2}\right)\cos\left(\frac{\alpha\tL}{2}\right)+\alpha\cosh\left(\frac{t_n^+\tL}{2}\right)\sin\left(\frac{\alpha\tL}{2}\right)\right\}e^{-v_n^+\tz_\p},\\
\label{a2}
G_{\rm c}(\tz_\p,\te)&=&2\sum_{n=0}^\infty\frac{b_n^+}{t{_n^+}^2+\alpha^2}\left\{\alpha\sinh\left(\frac{t_n^+\tL}{2}\right)\cos\left(\frac{\alpha\tL}{2}\right)-t_n^+\cosh\left(\frac{t_n^+\tL}{2}\right)\sin\left(\frac{\alpha\tL}{2}\right)\right\}e^{-v_n^+\tz_\p},\\
\label{a3}
F(\tz_\p,\te)&=&2\sideset{}{'}\sum_{n,m\geq0}b_n^+b_m^-\;\left[T\left(t_n^+,t_m^-\right)H\left(\pi/2-\te\right)+T\left(t_m^-,t_n^+\right)H\left(\te-\pi/2\right)\right]\;e^{-(v_n^++v_m^-)\tz_\p}.
\eea
In Eqs.~(\ref{a1})-(\ref{a3}), we introduced the auxiliary functions $b_0^\pm=u\pm1$, $b^\pm_{n>0}=\pm2\gamma^{2n}$, $v_n^\pm=2n\pm u$, $t_n^{\pm}=v_n^{\pm}\ct$, and $\alpha=\sqrt{u^2-1}\sin\te\cos\phi_k$. Eq.~(\ref{a3}) includes as well the function  
\bea
\label{a4}
T\left(t_n^+,t_m^-\right)&=&\frac{e^{-(t_n^+-t_m^-)\frac{\tL}{2}}}{\left(t{_n^+}^2+\alpha^2\right)\left(t{_m^-}^2+\alpha^2\right)}\left\{-\left(t_n^+t_m^-+\alpha^2\right)\cos(\alpha\tL)+\alpha(t_m^--t_n^+)\sin(\alpha\tL)\right\}\nonumber\\
&&+\frac{t_n^+e^{(t_n^++t_m^-)\frac{\tL}{2}}}{\left(t{_n^+}^2+\alpha^2\right)(t_n^++t_m^-)}+\frac{t_m^-e^{-(t_n^++t_m^-)\frac{\tL}{2}}}{\left(t{_m^-}^2+\alpha^2\right)(t_n^++t_m^-)}.
\eea
In Eq.~(\ref{a3}), the prime above the sum sign indicates that the term associated with the indices $n=m=0$ should be excluded from the summation.

\section{Auxiliary Function $R(\tz_\p,\te)$ of Eq.~(\ref{pm1l2})}
\label{ap2}

We report below the auxiliary function $R(\tz_\p)$ of Eq.~(\ref{pm1l2}).
\bea\label{R}
R(\tz_\p,\te)&=&S(u)\left[{\rm Arcth\left(\gamma e^{-\tz_-}\right)}-{\rm Arcth\left(\gamma e^{-\tz_+}\right)}\right]\\
&&+\frac{\gamma}{2}\td\left(1+u^{-1}\right)\left\{e^{-(2u+1)\tz_-}\Phi\left(\gamma^2e^{-2\tz_-},1,u+\frac{1}{2}\right)-e^{-(2u+1)\tz_+}\Phi\left(\gamma^2e^{-2\tz_+},1,u+\frac{1}{2}\right)\right\}\nonumber\\
&&+\td\left\{\gamma^{-2u}\left[\mb\left(\gamma^2e^{-2\tz_-},u+\frac{5}{2},-1\right)-\mb\left(\gamma^2e^{-2\tz_+},u+\frac{5}{2},-1\right)\right]\right.\nonumber\\
&&\hspace{8mm}\left.+\gamma^3\left[e^{-(2u+3)\tz_-}\Phi\left(\gamma^2e^{-2\tz_-},1,u+\frac{3}{2}\right)-e^{-(2u+3)\tz_+}\Phi\left(\gamma^2e^{-2\tz_+},1,u+\frac{3}{2}\right)\right]\right\}\nonumber\\
&&-\mb\left(\gamma^2e^{-2\tz_-},\frac{5}{2},-1\right)+\mb\left(\gamma^2e^{-2\tz_+},\frac{5}{2},-1\right)-\gamma^3\left[e^{-3\tz_-}\Phi\left(\gamma^2e^{-2\tz_-},1,\frac{3}{2}\right)-e^{-3\tz_+}\Phi\left(\gamma^2e^{-2\tz_+},1,\frac{3}{2}\right)\right].\nonumber
\eea

In Eq.~(\ref{R}), we defined the function
\be
S(u)=\frac{2+s^2}{s\sqrt{1+s^2}}-\td\left(u^{-1}+2u+\frac{2+3s^2}{s\sqrt{1+s^2}}\right)-1
\ee
together with the Lerch transcendent function $\Phi(x,n,a)$ and the incomplete Beta function $\mb(x,a,b)$ defined as~\cite{math}
\be
\Phi(x,n,a)=\sum_{i=0}^{\infty}\frac{x^i}{(i+a)^n};\hspace{1cm}\mb(x,a,b)=\int_0^x\md t\;t^{a-1}(1-t)^{b-1}.
\ee
\end{widetext}

\section{Effect of the Polyelectrolyte Rotations on Polyelectrolyte-Membrane Interactions in Counterion Liquids}
\label{pi}

In this appendix, we show that in counterion-only liquids, polyelectrolyte-membrane interactions are weakly altered by the polyelectrolyte orientation. First, by inserting the MF potential~(\ref{co1}) into Eq.~(\ref{pm1l}) and carrying out the integral, the MF component of the direct polyelectrolyte-membrane interaction energy follows as
\bea
\label{co8}
\beta\Omega^{(0)}_{\rm pm}(\bzp,\te)&=&-\frac{2\mu\tau}{q\cos\te}\left\{\left(1+\bz_+\right)\ln\left(1+\bz_+\right)\right.\\
&&\hspace{1.7cm}\left.-\left(1+\bz_-\right)\ln\left(1+\bz_-\right)\right\},\nonumber
\eea
with the rescaled coordinates of the polyelectrolyte edges
\be\label{co9}
\bz_\pm=\bzp\pm\frac{\bL}{2}\ct,
\ee
and the dimensionless polyelectrolyte distance $\bzp=z_\p/\mu$ and length $\bL=L/\mu$. 

\begin{figure*}
\includegraphics[width=.6\linewidth]{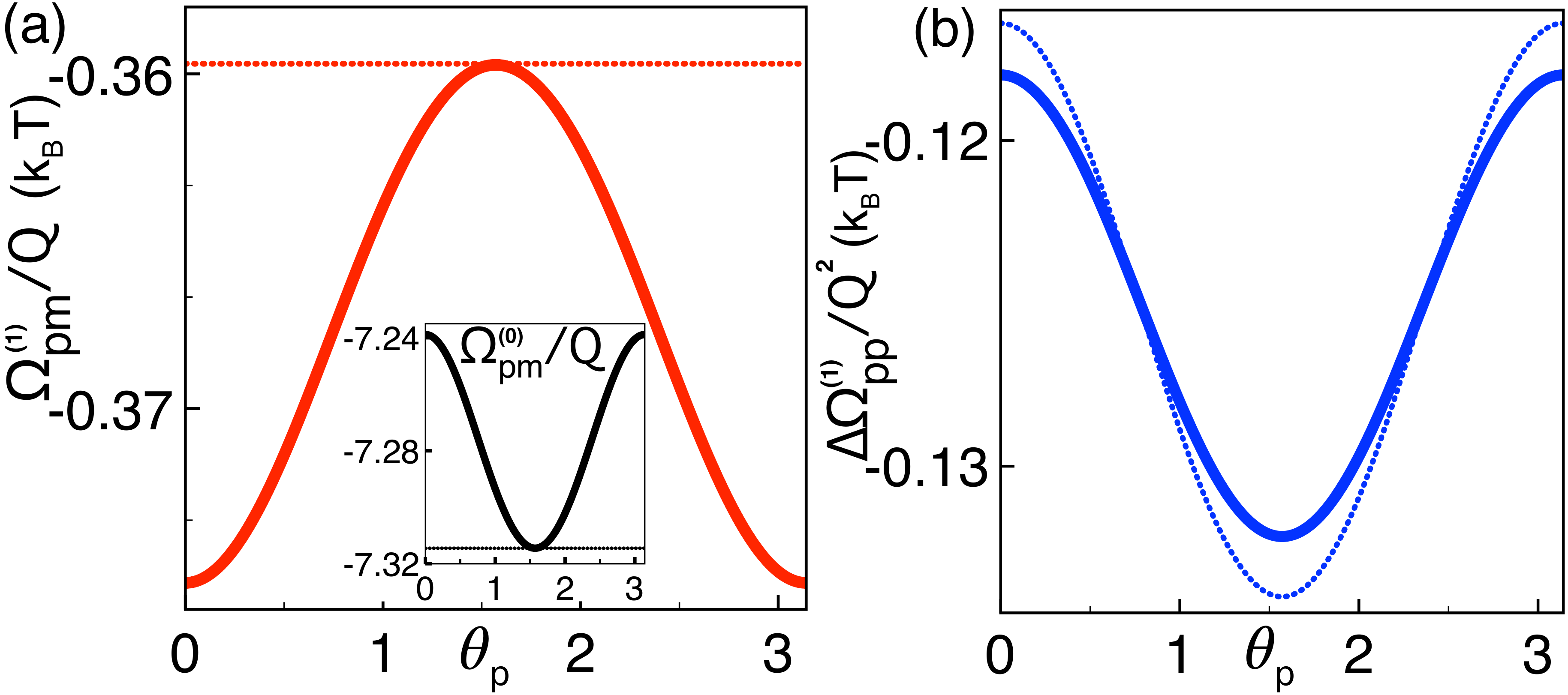}
\caption{(Color online)  (a) MF interaction potential~(\ref{co8}) (inset) and its 1l correction~(\ref{co12}) (main plot), and (b) polyelectrolyte self-energy~(\ref{co16}) versus the orientational angle of dsDNA in monovalent counterions (q=1). The polyelectrolyte distance is $z_{\rm p}=L$, length $L=3$ nm, and the membrane charge is $\sigma_{\rm m}=1.0$ $e/\mbox{nm}^2$. The dotted curves display the large distance limit of the polyelectrolyte potentials (see the main text).}
\label{figAp}
\end{figure*}

The inset of Fig.~\ref{figAp}(a) shows that despite the strong membrane charge and the close polyelectrolyte distance (see the caption), the orientation of the polyelectrolyte from $\te=0$ to $\te=\pi/2$ weakly modifies the MF interaction potential~(\ref{co8}). To elucidate this point, we consider the large distance regime $z_{\rm p}\gg L|\ct|/2$ where the MF interaction potential~(\ref{co9}) takes the asymptotic form
\be
\label{co9II}
\beta\Omega^{(0)}_{\rm pm}(\bzp,\te)\approx-\frac{2Q}{q}\left[1+\ln\left(1+\bzp\right)\right]
\ee
where $Q=L\tau$. Eq.~(\ref{co9II}) is indeed independent of the orientational angle $\te$ (see also the horizontal curve in the inset of Fig.~\ref{figAp}(a)). This is in contrast with the finite salt system where the large distance limit of the MF interaction potential in Eq.~(\ref{12}) was shown to depend strongly on the polyelectrolyte angle. 

In the salt-free system, the weak orientational dependence of polyelectrolyte-membrane interactions stems from their long range induced by the absence of salt-screening. This decreases the variation of the electrostatic force from the lower portion ($l<L/2$) to the upper portion of the molecule ($l>L/2$), reducing the multipolar component of the polyelectrolyte-membrane interactions susceptible to the orientation of the molecule.

We compute now the 1l correction to the MF interaction potential~(\ref{co8}). By substituting the average potential correction~(\ref{co7}) into Eq.~(\ref{pm1l}), one finds
\be
\label{co12}
\beta\Omega^{(1)}_{\rm pm}(\bzp,\te)=\frac{q\ell_B\tau}{4}\int_0^\infty\frac{\mathrm{d}\rk}{\rk^2}\chi(\bzp,\te),
\ee
with the auxiliary function
\begin{widetext}
\bea
\label{aco6}
\chi(\bzp,\te)&=&\frac{\bL}{(\bz_++1)(\bz_-+1)}+\frac{2}{\ct}\left[\Delta(\rk+1)-1\right]\ln\left(\frac{\bz_++1}{\bz_-+1}\right)\nonumber\\
&&-\frac{2\Delta\rk}{\ct}e^{2\rk}\left\{\frac{e^{-2\rk(1+\bz_-)}}{2\rk(1+\bz_-)}-\frac{e^{-2\rk(1+\bz_+)}}{2\rk(1+\bz_+)}+{\rm Ei}\left[-2\rk(1+\bz_-)\right]-{\rm Ei}\left[-2\rk(1+\bz_+)\right]\right\}.
\eea
\end{widetext}
Fig.~\ref{figAp}(a) shows that the potential~(\ref{co12}) minimized at $\te=0$ and $\pi$ favors the perpendicular polyelectrolyte orientation (solid curve). This said, one notes again a perturbative variation of Eq.~(\ref{co12}) by the polyelectrolyte rotation. At large separation distances $z_{\rm p}\gg L|\ct|/2$,  Eq.~(\ref{aco6}) simplifies to 
\bea
\label{aco7}
\chi(\bzp,\te)&\approx&\frac{\bL}{\left(1+\bzp\right)^2}\left\{-\Delta e^{-2\rk\bzp}+1\right.\\
&&\hspace{1.7cm}\left.+2\left(1+\bzp\right)\left[-1+\Delta\left(\rk+1\right)\right]\right\}.\nonumber
\eea
Evaluating the integral~(\ref{co12}) with Eq.~(\ref{aco7}), one gets 
\bea
\label{co19}
\beta\Omega^{(1)}_{\rm pm}(\bzp,\te)&\approx&\frac{\Xi_{\rm c} Q}{8q\left(1+\bzp\right)^2}\left\{4\bzp-4\pi e^{\bzp}\sin(\bzp)\right.\\
&&\hspace{1.cm}-\left[4\gamma_{\rm e}+\pi+\ln\left(4\bzp^4\right)\right](1+\bzp)\nonumber\\
&&\hspace{1.cm}\left.+4\;\mathrm{Re}\left[e^{(1+i)\bzp}\mathrm{Ei}\left[-(1+i)\bzp\right]\right]\right\}.\nonumber
\eea
Eq.~(\ref{co19}) reported in Fig.~\ref{figAp}(a) by the dotted curve is indeed independent of the angle $\te$. This is again in contrast with the finite salt system where the 1l potential correction~(\ref{sas3})  was shown to depend strongly on the polyelectrolyte angle (see Fig.~\ref{fig4}(c)). 

Finally, we consider the polyelectrolyte self-energy. Inserting the Green's function in Eq.~(\ref{co5}) into Eq.~(\ref{9II}), after lengthy algebra, the self energy follows as
\begin{widetext}
\be
\label{co10}
\beta\Delta\Omega_{\rm pp}^{(1)}(\bzp,\te)=\frac{\mu\ell_B\tau^2}{2}\int_0^{2\pi}\frac{\mathrm{d}\phi_k}{2\pi}\int_0^\infty\frac{\mathrm{d}\rk}{\rk^2}\left[I_1(\bzp,\te)+\Delta I_2(\bzp,\te)\right],
\ee
with the auxiliary functions 
\bea\label{aco1}
I_1(\bzp,\te)&=&2\;{\rm Re}\left[J(x_+,x_-,v_+)-J_{\rm b}(v_+)\right]H\left(\pi/2-\te\right)+2\;{\rm Re}\left[J(x_-,x_+,v_+)-J_{\rm b}(-v_+)\right]H\left(\te-\pi/2\right),\\
\label{aco3}
I_2(\bzp,\te)&=&e^{2\rk}\left|\frac{1}{v_-}\left(e^{-v_-x_-}-e^{-v_-x_+}\right)+\frac{1}{\ct}\left[{\rm Ei}(-v_-x_+)-{\rm Ei}(-v_-x_-)\right]\right|^2,
\eea
where we used the auxiliary parameters $x_\pm=\rk(\bz_\pm+1)/\ct$ and $v_\pm=\ct\pm it$, with $t=\sin\te\cos\phi_k$, and the functions 
\bea
\label{aco4}
J(x_+,x_-,v_+)&=&\frac{1}{v_+}\left(x_+-x_-\right)-\frac{1}{v_+\ct}\left\{e^{-v_+x_-}{\rm Ei}(v_+x_-)-e^{-v_+x_+}{\rm Ei}(v_+x_+)\right\}\nonumber\\
&&+\left[\frac{{\rm Ei}(v_+x_-)}{\ct}-\frac{e^{v_+x_-}}{v_+}\right]\left\{\frac{1}{v_+}\left(e^{-v_+x_-}-e^{-v_+x_+}\right)+\frac{1}{\ct}\left[{\rm Ei}(-v_+x_+)-{\rm Ei}(-v_+x_-)\right]\right\}\nonumber\\
&&+\frac{i\pi}{\cos^2\te}{\rm sgn}\left[\cos\left(\te\right)\cos\left(\phi_k\right)\right]\left[\Gamma(0,v_+x_+)-\Gamma(0,v_+x_-)\right]\nonumber\\
&&-\frac{1}{\cos^2\te}\left\{\MeijerG[\Bigg]{3}{1}{2}{3}{0,1}{0,0,0}{-v_+x_+}-\MeijerG[\Bigg]{3}{1}{2}{3}{0,1}{0,0,0}{-v_+x_-}\right\},\\
\label{aco5}
J_{\rm b}(v_+)&=&\frac{1}{v_+}\left\{\rk\bL-\frac{1}{v_+}\left(1-e^{-v_+\rk\bL}\right)\right\}.
\eea
In Eq.~(\ref{aco3}) and~(\ref{aco4}), we used the exponential integral function ${\rm Ei}(x)$, the sign function ${\rm sgn}(x)$, the incomplete gamma function $\Gamma(0,x)$, and the Meijer-G function $\MeijerG*{m}{n}{p}{q}{a_1, \dots, a_p}{b_1, \dots, b_q}{x}$~\cite{math}.  As the integrand of Eq.~(\ref{co10}) includes special functions with complex arguments, the numerical evaluation of the double integral with sufficient precision turned out to be a very difficult task. Thus, we focus on the large distance regime $z_{\rm p}+\mu\gg L|\ct|/2$ where the coefficients in Eqs.~(\ref{aco1}) and~(\ref{aco3}) converge to
\bea\label{co13}
I_1&\approx&\frac{2}{\left(\cos^2\te+t^2\right)^3\left(1+\bzp\right)^2\rk^2}\left\{-t^4-2t^2\left(3+\rk\bL\ct\right)\cos^2\te-\left(-3+2\rk\bL\ct\right)\cos^4\te\right.\nonumber\\
&&\hspace{4.4cm}+\left[t^4\left(1+\rk\bL\ct\right)+6t^2\cos^2\te-\left(3+\rk\bL\ct\right)\cos^4\te\right]e^{-\rk\bL\ct}\cos\left(t\rk\bL\right)\nonumber\\
&&\hspace{4.4cm}\left.+2t\cos^2\te\left[\left(4+\rk\bL\ct\right)\ct+t^2\rk\bL\right]e^{-\rk\bL\ct}\sin\left(t\rk\bL\right)\right\},\\
\label{co14}
I_2&\approx&\frac{2}{\cos^2\te+t^2}\left[\cosh\left(\rk\bL\ct\right)-\cos\left(t\rk\bL\right)\right]\left[1+\frac{1}{\rk\left(1+\bzp\right)}\right]^2e^{-2\rk\bzp}.
\eea
\end{widetext}
In order to evaluate analytically the self-energy in Eq.~(\ref{co10}), one needs to approximate the expression in the first bracket of Eq.~(\ref{co14}) by its Taylor-expansion of order $O\left[\left(\rk\bL\right)^4\right]$. This yields
\be
\label{co15}
I_2\approx\left[1+\frac{\left(\rk\bL\right)^2}{12}\left(\cos^2\te-t^2\right)\right]\left[\rk\bL+\frac{\bL}{\left(1+\bzp\right)}\right]^2e^{-2\rk\bzp}.
\ee
Within these approximations, one can carry-out the double integral in Eq.~(\ref{co10}) to get 
\be\label{co16}
\beta\Delta\Omega_{\rm pp}^{(1)}(\bzp,\te)\approx\frac{\Xi_{\rm c} Q^2}{384q^2}\psi(\bzp,\te),
\ee
with the dimensionless self-energy
\bea
\label{co17}
\psi(\bzp,\te)&=&2\;\mathrm{Re}\left\{\left\{\bL^2\left[1+3\cos(2\te)\right]-96i\right\}e^{(1+i)\bzp}\right.\nonumber\\
&&\hspace{6mm}\times\left.\left\{i\pi+\mathrm{Ei}\left[(-1-i)\bzp\right]\right\}\left(\frac{\bzp+i}{\bzp+1}\right)^2\right\}\nonumber\\
&&+\frac{1}{\bzp\left(1+\bzp\right)^2}\left\{64\bL\bzp-384\bzp^2\right.\nonumber\\
&&\hspace{1.cm}\left.+\bL^2\left[1+3\cos(2\te)\right]\left(\bzp+1\right)^2\right\}
\eea
where we used the exponential integral function ${\rm Ei}(x)$~\cite{math}. Eq.~(\ref{co16}) is plotted in Fig.~\ref{figAp}(b) against the polyelectrolyte angle (solid curve). The figure shows that the polyelectrolyte rotation from $\te=0$ to $\te=\pi/2$ alters the self-energy only by about 10\%. To gain analytical insight, we consider the strict large distance regime $\bzp\gg1$ where Eq.~(\ref{co17}) takes the asymptotic form
\bea
\label{co18}
\beta\Delta\Omega_{\rm pp}^{(1)}(\bzp,\te)&\approx&\frac{\Xi_{\rm c} Q^2}{q^2}\left\{-\frac{3}{4\bzp}+\frac{2\bL+9}{12\bzp^2}\right.\\
&&\left.\hspace{-1.2cm}+\frac{1}{384\bzp^3}\left[5\bL^2-128\bL+15\bL^2\cos(2\te)-288\right]\right\}\nonumber
\eea
reported in Fig.~\ref{figAp}(b) by the dotted curve. The weak angular dependence of the self-energy stems from the fact that only the third order perturbative term of the asymptotic expansion~(\ref{co18}) depends on the angle $\te$. Since we have shown that the MF and 1l interaction energy components $\Omega^{(i)}_{\rm pm}(\bzp,\te)$ behave similarly, we can conclude that in the counterion-only liquid, the total 1l grand potential~(\ref{gr1l}) is weakly affected by the polyelectrolyte orientation.

\end{document}